%% file: S5KnownPulsarPaper.tex
\documentclass[preprint,12pt]{aastex}

\usepackage{amsmath}
\usepackage{color}

\newcommand{\ee}[1]{\!\times\!10^{#1}}
\newcommand{\gws}{gravitational waves~}
\newcommand{\gw}{gravitational wave~}

\begin{document}


\title{Searches for gravitational waves from known pulsars with S5 LIGO data}
\shorttitle{Gravitational waves from pulsars}

\input{authorlist.tex}

\begin{abstract}
We present a search for gravitational waves from 116 known millisecond and young
pulsars using data from the fifth science run of the LIGO detectors. For this
search ephemerides overlapping the run period were obtained for all pulsars
using radio and X-ray observations. We demonstrate an updated search method
that allows for small uncertainties in the pulsar phase parameters to be
included in the search. We report no signal detection from any of the targets
and therefore interpret our results as upper limits on the \gw signal strength.
The most interesting limits are those for young pulsars. We present updated
limits on gravitational radiation from the Crab pulsar, where the measured limit
is now a factor of seven below the spin-down limit. This limits the power
radiated via \gws to be less than $\sim$2\% of the available spin-down power.
For the X-ray pulsar J0537$-$6910 we reach the spin-down limit under the
assumption that any \gw signal from it stays phase locked to the X-ray pulses
over timing glitches, and for pulsars J1913+1011 and J1952+3252 we are only a
factor of a few above the spin-down limit. Of the recycled millisecond pulsars
several of the measured upper limits are only about an order of magnitude above
their spin-down limits. For these our best (lowest) upper limit on \gw amplitude
is $2.3\ee{-26}$ for J1603$-$7202 and our best (lowest) limit on the inferred
pulsar ellipticity is $7.0\ee{-8}$ for J2124$-$3358.
\end{abstract}


\maketitle


\section{Introduction}\label{sec:intro}
Within our Galaxy some of the best targets for \gw searches in the sensitive
frequency band of current interferometric \gw detectors ($\sim$40--2000\,Hz)
are millisecond and young pulsars. There are currently just over 200 known
pulsars with spin frequencies greater than 20\,Hz, which therefore are within this
band. In this paper we describe the latest results from the ongoing search for
\gws from these known pulsars using data from the Laser Interferometric
Gravitational-Wave Observatory (LIGO). As this search looks for objects with
known positions and spin-evolutions it can use long time spans of data in a
fully coherent way to dig deeply into the detector noise. Here we use data from
the entire two-year run of the three LIGO detectors, entitled Science Run 5
(S5), during which the detectors reached their design sensitivities
\citep{Abbott:2007b}. This run started on 2005 November 4 and ended on 2007
October 1. The detectors (the 4\,km and 2\,km detectors at LIGO Hanford
Observatory, H1 and H2, and the 4\,km detector at the LIGO Livingston
Observatory, L1) had duty factors of 78\% for H1, 79\% for H2, and 66\% for L1.
The GEO\,600 detector also participated in S5 \citep{Grote:2008}, but at lower
sensitivities that meant it was not able to enhance this search. The Virgo
detector also had data overlapping with S5 during Virgo Science Run 1 (VSR1)
\citep{Acernese:2008}. However this was also generally at a lower sensitivity
than the LIGO detectors and had an observation time of only about 4 months,
meaning that no significant sensitivity improvements could be made by including
this data. Due to its multi-stage seismic isolation system Virgo does have
better sensitivity than the LIGO detectors below about 40\,Hz, opening the
possibility of searching for more young pulsars, including the Vela pulsar.
These lower frequency searches will be explored more in the future.

This search assumes that the pulsars are triaxial stars emitting \gws at
precisely twice their observed spin frequencies, i.e.\ the emission mechanism is
an $\ell=m=2$ quadrupole, and that \gws are phase-locked with the
electromagnetic signal. We use the so-called spin-down limit on strain tensor
amplitude $h_0^{\rm sd}$ as a sensitivity target for each pulsar in our
analysis. This can be calculated, by assuming that the observed spin-down rate
of a pulsar is entirely due to energy loss through gravitational radiation from
an $\ell=m=2$ quadrupole, as
\begin{equation}
h_0^{\rm sd} = 8.06\ee{-19}I_{38}r^{-1}_{\rm kpc}(|\dot{\nu}|/\nu)^{1/2},
\end{equation}
where $I_{38}$ is the pulsar's principal moment of inertia ($I_{zz}$) in units
of $10^{38}$\,kg\,m$^2$, $r_\mathrm{kpc}$ is the pulsar distance in kpc, $\nu$
is the spin-frequency in Hz, and $\dot{\nu}$ is the spin-down rate in
Hz\,s$^{-1}$. Due to uncertainties in $I_{zz}$ and $r$, $h_0^{\rm sd}$ is
typically uncertain by about a factor 2. Part of this is due to the uncertainty
in $I_{zz}$ which, though predicted to lie roughly in the range
1--$3\ee{38}$\,kg\,m$^2$, has not been measured for any neutron star; and the
best (though still uncertain) prospect is star A of the double pulsar system
J0737-3039 with 20 years' more observation \citep{Kramer:2009}. Distance
estimates based on dispersion measure can also be wrong by a factor 2--3, as
confirmed by recent parallax observations of the double pulsar
\citep{Deller:2009}. For pulsars with measured braking indices,
$n=\nu\ddot{\nu}/\dot{\nu}^2$, the assumption that spin-down is dominated by \gw
emission is known to be false (the braking index for quadrupolar \gw emission
should be 5, but all measured $n$'s are less than 3) and a stricter indirect
limit on \gw emission can be set. A phenomenological investigation of some young
pulsars \citep{Palomba:2000} indicates that this limit is lower than $h_0^{\rm
sd}$ by a factor 2.5 or more, depending on the pulsar. See \citet{Abbott:2007a}
and \citet{Abbott:2008} for more discussion of the uncertainties in indirect
limits. Recycled millisecond pulsars have intrinsically small spin-downs, so
for the majority of pulsars in our search these spin-down limits will be well
below our current sensitivities, making detection unlikely. However, our search
also covers four young pulsars with large spin-down luminosities, and for these
we can potentially beat or reach their spin-down limits using current data.

The LIGO band covers the fastest (highest-$\nu$) known pulsars, and the
quadrupole formula for strain tensor amplitude
\begin{equation}
h_0 = 4.2\times10^{-26} \nu_{100}^2 I_{38} \varepsilon_{-6} r_\mathrm{kpc}^{-1}
\end{equation}
indicates that these pulsars are the best \gw emitters for a given equatorial
ellipticity $\varepsilon = (I_{xx} - I_{yy}) /I_{zz}$ (here $\nu_{100} =
\nu/(100~\mathrm{Hz})$ and $\varepsilon_{-6} = \varepsilon/10^{-6}$). The
pulsars with high spin-downs are almost all less than $\sim10^4$ years old.
Usually this is interpreted as greater electromagnetic activity (including
particle winds) in younger objects, but it could also mean that they are more
active in \gw emission. This is plausible on theoretical grounds too. Strong
internal magnetic fields may cause significant ellipticities \citep{Cutler:2002}
which would then decay as the field decays or otherwise changes
\citep{GoldreichReisenegger}. The initial crust may be asymmetric if it forms on
a time scale on which the neutron star is still perturbed by its violent
formation and aftermath, including a possible lengthy perturbation due to the
fluid $r$-modes \citep{Lindblom:2000gu, Wu:2000qy}, and asymmetries may slowly
relax due to mechanisms such as viscoelastic creep. Also the fluid $r$-modes may
remain unstable to \gw emission for up to a few thousand years after the neutron
star's birth, depending on its composition, viscosity, and initial spin
frequency \citep{Owen:1998xg, Bondarescu:2008qx}. Such $r$-modes are expected to
have a \gw frequency about 4/3 the spin frequency. However, we do not report on
$r$-mode searches in this paper.

\subsection{Previous analyses}
The first search for \gws from a known pulsar using LIGO and GEO\,600 data came
from the first science run (S1) in 2002 September. This targeted just one
pulsar in the approximately one weeks worth of data -- the then fastest known
pulsar J1939+2134 \citep{Abbott:2004}. Data from LIGO's second science run (S2),
which spanned from 2003 February to 2003 April, was used to search for 28
isolated pulsars (i.e.\ those not in binary systems) \citep{Abbott:2005}. The
last search for \gws from multiple known pulsars using LIGO data combined data
from the third and fourth science runs and had 78 targets, including isolated
pulsars and those in binary systems \citep{Abbott:2007a}. The best (lowest),
95\% degree-of-belief, upper limit on \gw amplitude obtained from the search was
$h_0^{95\%} = 2.6\ee{-25}$ for J1603$-$7202, and the best (smallest) limit on
ellipticity was just under $10^{-6}$ for J2124$-$3358. The data run used in this
paper is almost an order of magnitude longer, and has a best strain noise
amplitude around a factor of two smaller, than that used in the best previous 
search. 

We have also previously searched the first nine months of S5 data for a signal
from the Crab pulsar \citep{Abbott:2008}. That analysis used two methods to
search for a signal: one in which the signal was assumed to be precisely
phase-locked with the electromagnetic signal, and another which searched a small
range of frequencies and frequency derivatives around the electromagnetic
parameters. The time span of data analysed was dictated by a timing glitch in
the pulsar on 2006 August 23, which was used as the end point of the analysis.
In that search the spin-down limit for the Crab pulsar was beaten for the first
time (indeed it was the first time a spin-down limit had been reached for any
pulsar), with a best limit of $h_0^{95\%} = 2.7\ee{-25}$, or slightly below
one-fifth of the spin-down limit. This allowed the total power radiated in \gws
to be constrained to less than 4\% of the spin-down power. We have since
discovered an error in the signal template used for the search
\citep{CrabErratum}. We have re-analysed the data and find a new upper limit
based on the early S5 data alone at the higher value shown in
Table~\ref{tab:crabresults}, along with the smaller upper limit based on the
full S5 data.

For this analysis we have approximately 525 days of H1 data, 532 days of H2 data
and 437 days of L1 data. This is using all data flagged as {\it science mode}
during the run (i.e.\ taken when the detector is locked in its operating
condition on the dark fringe of the interference pattern, and relatively
stable), except data one minute prior to loss of lock, during which time it is
often seen to become more noisy.

\subsection{Electromagnetic observations}\label{sec:observations}
The radio pulsar parameters used for our searches are based on ongoing radio 
pulsar monitoring programs, using data from the Jodrell Bank Observatory (JBO),
the NRAO 100\,m Green Bank Telescope (GBT) and the Parkes radio telescope of the
Australia Telescope National Facility. We used radio data coincident with the S5
run as these would reliably represent the pulsars' actual phase evolution during
our searches. We obtained data for 44 pulsars from JBO (including the Crab
pulsar ephemeris, \citet{CrabEphemerisPaper, CrabEphemeris}), 39 pulsars within
the Terzan 5 and M28 globular clusters from GBT, and 47 from Parkes, including
pulsars timed as part of the Parkes Pulsar Timing Array~\citep{Manchester:2008}.
For 15 of these pulsars there were observations from more than one site, making
a total of 115 radio pulsars in the analysis (see Table~\ref{tab:results} for
list of the pulsars, including the observatory and time span of the
observations). For the pulsars observed at JBO and Parkes we have obtained
parameters fit to data overlapping with the entire S5 run. For the majority of
pulsars observed at GBT the parameters have been fit to data overlapping
approximately the first quarter of S5.

Pulsars generally exhibit timing noise on long time scales. Over tens of years
this can cause correlations in the pulse time of arrivals which can give
systematic errors in the parameter fits produced, by the standard pulsar timing
package TEMPO\footnote{\url{http://www.atnf.csiro.au/research/pulsar/tempo/}},
of order 2--10 times the regular errors that TEMPO assigns to each parameter
\citep{Verbiest:2008}, depending on the amplitude of the noise. For our pulsars,
with relatively short observation periods of around two years, the long-term
timing noise variations should be largely folded in to the parameter fitting,
leaving approximately white uncorrelated residuals. Also millisecond pulsars, in
general, have intrinsically low levels of timing noise, showing comparatively
white residuals. This should mean that the errors produced by TEMPO are
approximately the true 1$\sigma$ errors on the fitted values.

The regular pulse timing observations of the Crab pulsar
\citep{CrabEphemerisPaper, CrabEphemeris} indicate that the 2006 August 23
glitch was the only glitch during the S5 run. One other radio pulsar,
J1952+3252, was observed to glitch during the run (see \S\ref{subsec:J1952}.)
Independent ephemerides are available before and after each glitch.

We include one pulsar in our analysis that is not observed as a radio pulsar.
This is PSR\,J0537$-$6910 in the Large Magellanic Cloud, for which only X-ray
timings currently exist. Data for this source come from dedicated time on the
Rossi X-ray Timing Explorer (RXTE) \citep{Middleditch:2006}, giving ephemerides
covering the whole of S5. These ephemerides comprise seven inter-glitch
segments, each of which produces phase-stable timing solutions. The segments are
separated by times when the pulsar was observed to glitch. Due to the
complexity of the pulsar behaviour near glitches, which is not reflected in
the simple model used to predict pulse times of arrival, sometimes up to $\sim
30$ days around them are not covered by the ephemerides.

\section{Gravitational wave search method}\label{sec:method}
The details of the search method are discussed in \citet{Dupuis:2005} and
\citet{Abbott:2007a}, but we will briefly review them here. Data from the \gw
detectors are heterodyned using twice the known electromagnetic phase evolution
of each pulsar, which removes this rapidly varying component of the signal,
leaving only the daily varying amplitude modulation caused by each detector's
antenna response. Once heterodyned the (now complex) data are low-pass
filtered at 0.25\,Hz, and then heavily down-sampled, by averaging, from the
original sample rate of 16\,384\,Hz to 1/60\,Hz. Using these down-sampled data
($B_k$, where $k$ represents the $k^{\rm th}$ sample) we perform parameter
estimation over the signal model $y_k(\boldsymbol{a})$ given the unknown signal
parameters $\boldsymbol{a}$. This is done by calculating the posterior
probability distribution \citep{Abbott:2007a}
\begin{equation}
p(\boldsymbol{a}|\{B_k\}) \propto \prod_j^M \left(\sum_{k}^{n}
(\Re\{B_k\}-\Re\{y_k(\boldsymbol{a})
\})^2 + (\Im\{B_k\}-\Im\{y_k(\boldsymbol{a})\})^2\right)^{-m_j} \times
p(\boldsymbol{a}),
\end{equation}
where the first term on the right hand side is the likelihood (marginalised over
the data variance, giving a Student's-t-like distribution), $p(\boldsymbol{a})$
is the prior distribution for $\boldsymbol{a}$, $M$ is the number of data
segments into which the $B_k$s have been cut (we assume stationarity of the data
during each segment), $m_j$ is the number of data points in the $j$th
segment (with a maximum value of 30, i.e.\ we only assume stationarity for
periods less than, or equal to, 30 minutes in length), and
$n=\sum_{j=1}^{j}m_j$. The assumption of Gaussianity and stationarity of the
segments holds well for this analysis (see \S4.5 of \citet{Dupuis:2004} for
examples of $\chi^2$ and Kolmogorov-Smirnov tests performed to assess these in
previous analyses).

We have previously \citep{Abbott:2004,Abbott:2005,Abbott:2007a} performed
parameter estimation over the four unknown \gw signal parameters of amplitude
$h_0$, initial phase $\phi_0$, cosine of the orientation angle $\cos{\iota}$,
and polarisation angle $\psi$, giving $\boldsymbol{a} = \{h_0, \phi_0,
\cos{\iota}, \psi\}$. Priors on each parameter are set to be uniform over their
allowed ranges, with the upper end of the range for $h_0$ set empirically from 
the noise level of the data.  We choose a uniform prior on $h_0$ for consistency
with our previous analyses
\citep{Dupuis:2005,Abbott:2004,Abbott:2005,Abbott:2007a} and to facilitate
straightforward comparison of sensitivity. Extensive trials with software
injections have shown this to be a very reasonable choice, returning a
conservative (i.e.~high) upper limit consistent with the data and any possible
signal.

Using a uniformly spaced grid on this four-dimensional parameter space the
posterior is calculated at each point. To obtain a posterior for each individual
parameter we marginalise over the three others. Using the marginalised posterior
on $h_0$ we can set an upper limit by calculating the value that, integrating up
from zero, bounds the required cumulative probability (which we have taken as
95\%). We also combine the data from multiple detectors to give a {\it joint}
posterior. To do this we simply take the product of the likelihoods for each
detector and multiply this joint likelihood by the prior. This is possible due
to the phase coherence between detectors. Again we can marginalise to produce
posteriors for individual parameters.

Below, in \S\ref{sec:mcmc}, we discuss exploring and expanding this parameter
space to more dimensions using a Markov chain Monte Carlo (MCMC) technique.

\subsection{MCMC parameter search}\label{sec:mcmc}
When high resolutions are needed it can be computationally time consuming to
calculate the posterior over an entire grid as described above, and redundant
areas of parameter space with very little probability are explored for a
disproportionately large amount of time. A more efficient way to carry out such
a search is with a Markov chain Monte Carlo (MCMC) technique, in which the
parameter space is explored more efficiently and without spending much time in
the areas with very low probability densities.

An MCMC integration explores the parameter space by stepping from one position
in parameter space to another, comparing the posterior probability of the two
points and using a simple algorithm to determine whether the step should be
accepted. If accepted it moves to that new position and repeats; if it is
rejected it stays at the current position and repeats. Each iteration of the
chain, whether it stays in the same position or not, is recorded and the amount
of time the chain spends in a particular part of parameter space is directly
proportional to the posterior probability density there. The new points are
drawn randomly from a specific proposal distribution, often given by a
multivariate Gaussian with a mean set as the current position, and a predefined
covariance. For an efficient MCMC the proposal distribution should reflect the
underlying posterior it is sampling, but any proposal (that does not explicitly
exclude the posterior), given enough time, will sample the posterior and
deliver an accurate result. We use the Metropolis-Hastings (MH) algorithm to set
the acceptance/rejection ratio. Given a current position $\boldsymbol{a}_i$ MH
accepts the new position $\boldsymbol{a}_{i+1}$ with probability
\begin{equation}\label{eq:mcmcmh}
\alpha(\boldsymbol{a}_{i+1}|\boldsymbol{a}_i) = {\rm min}\left(1,
\frac{p(\boldsymbol{a}_{i+1}|d)}{p(\boldsymbol{a}_i|d)}
\frac{q(\boldsymbol{a}_i|\boldsymbol{a}_{i+1})}{q(\boldsymbol{a}_{i+1}
|\boldsymbol{a}_i)} \right),
\end{equation}
where $p(\boldsymbol{a}|d)$ is the posterior value at $\boldsymbol{a}$ given
data $d$, and $q(\boldsymbol{a}|\boldsymbol{b})$ is the proposal distribution
defining how we choose position $\boldsymbol{a}$ given a current position
$\boldsymbol{b}$. In our case we have symmetric proposal distributions, so
$q(\boldsymbol{a}_{i+1}|\boldsymbol{a})/q(\boldsymbol{a}_i|\boldsymbol{a}_{i+1})
= 1$ and therefore only the ratio of the posteriors is needed.

A well-tuned MCMC will efficiently explore the parameter space and generate
chains that, in histogram form, give the marginalised posterior distribution for
each parameter. Defining a good set of proposal distributions for the parameters
in $\boldsymbol{a}$ has been done experimentally assuming that they are
uncorrelated and therefore have independent distributions. (There are in fact
correlations between the $h_0$ and $\cos{\iota}$ parameters and the $\phi_0$ and
$\psi$ parameters, but in our studies these do not significantly alter the
efficiency from assuming independent proposals.) The posterior distributions of
these parameters will also generally not be Gaussian, especially in low SNR
cases (which is the regime in which we expect to be), but a Gaussian proposal is
easiest to implement and again does not appear to significantly affect the chain
efficiency. We find that, for the angular parameters, Gaussian proposal
distributions with standard deviations of an eighth the allowed parameter range
(i.e.\ $\sigma_{\phi_0} = \pi/4$\,rad, $\sigma_{\cos{\iota}} = 1/4$ and
$\sigma_{\psi} = \pi/16$\,rad) provide a good exploration of the parameter space
(as determined from the ratio of accepted to rejected jumps in the chain) for
low SNR signals. We have performed many simulations comparing the output of the
MCMC and grid-based searches, both on simulated noise and simulated signals, and
both codes give results consistent to within a few percent. In these tests we
find that the computational speed of the MCMC code is about three times faster
than the grid-based code, although this can vary by tuning the codes.

An MCMC integration may take time to converge on the bulk of the probability
distribution to be sampled, especially if the chains start a long way in
parameter space from the majority of the posterior probability. Chains are
therefore allowed a {\it burn-in} phase, during which the positions in the chain
are not recorded. For low SNR signals, where the signal amplitude is close to
zero and the posteriors are reasonably broad, this burn-in time can be short. To
aid the convergence we use simulated annealing in which a temperature parameter
is used to flatten the posterior during burn-in to help the chain explore the
space more quickly. We do however use techniques to assess whether our chains
have converged (see \citet{Brooks:1998} for a good overview of convergence
assessment tests for MCMCs.) We use two such tests: the Geweke test is used on
individual chains to compare the means of two independent sections of the chain;
and the Gelman and Rubins test is used to compare the variances between and
within two or more separate chains. These tests are never absolute indicators of
convergence, so each chain also has to be examined manually. The
acceptance/rejection ratio of each chain is also looked at as another indicator
of convergence. For all our results, discussed in \S\ref{sec:analysis} and
\S\ref{sec:results}, we have run these tests on the output chains, and conclude
that all have converged.

\subsection{Adding phase parameters}\label{subsec:phaseparams}
The heterodyne phase is calculated using the parameters measured from
electromagnetic observations, which have associated errors. These errors could
mean that the heterodyne phase will be offset, and drift away from, the true
phase of the signal. Previously we have required that our data be heterodyned
with a phase that was known to match the true signal phase over the course of
the run to a few degrees. The criterion used to decide on whether to keep, or
discard, a pulsar from the analysis was that there was no more than a
30$^{\circ}$ drift of the electromagnetic phase from the signal phase over the
course of a data run \citep{Abbott:2007a} (i.e.\ if a signal was present the
phase drift would lead to a loss in SNR of less than about 15\%.) In
\citet{Abbott:2007a} this potential phase drift was calculated from the known
uncertainties of the heterodyne phase parameters, but without taking into
account the covariance between parameters, and as such was an over-conservative
estimate, risking the possibility that some pulsars were excluded from the
analysis unnecessarily.

Rather than just setting an exclusion criterion for pulsars based on the
potential phase mismatch (see \S\ref{sec:eval}) we can instead search over
the uncertainties in the phase parameters. This search can also be
consistent with the, now provided, covariances on the phase parameters. For
pulsars that have small mismatches over the run (that would have been included
in the previous analysis), the extra search space allows these small
uncertainties to be naturally folded (via marginalisation) into the final
posteriors on the four main \gw parameters and our eventual upper limit. For
pulsars with larger mismatches, which previously would have been excluded, this
extra search space allows us to keep them in the analysis and again fold the
phase parameter uncertainties into the final result.

We can incorporate the potential phase error into the search by including the
phase parameters, as an offset from the heterodyne phase parameters, in the
parameter estimation and marginalising over them. A pulsar with, for example,
associated errors on $\nu$, $\dot{\nu}$, right ascension, declination, proper
motion in both positional parameters and 5 orbital binary parameters would add
11 extra parameters to the search. An MCMC is a practical way that allows us to
search over these extra parameters. It also means that we make sure we cover
enough of the parameter space so as not to miss a signal with a slight phase
offset from the heterodyne values. Examples of MCMCs being used in a
similar context can be found in \citet{Umstatter:2004} and \citet{Veitch:2005}.
However both these examples attempted to explore far greater parameter ranges
than will be applied here. We do not attempt to use the MCMC as a parameter
estimation tool for these extra parameters (i.e.\ due to our expected low SNR
regime we would not expect to improve on the given uncertainties of the
parameters), but just as a way of making sure we fully cover the desired
parameter space, and fold the uncertainties into our final result without
excessive computational cost.

The number of extra parameters included in the search depends on how many
parameters were varied to fit the radio data. When fitting parameters with
the pulsar timing packages TEMPO, or TEMPO2~\citep{Hobbs:2006}, certain values
can be held fixed and others left free to vary, so errors will only be given on
those allowed to vary. The uncertainties on those parameters that are fit will
contain all the overall phase uncertainty.

\subsection{Setting up the MCMC}
In our search we have information on the likely range over which to explore each
parameter as given by the error from the fit (which we take as being the
1$\sigma$ value of a Gaussian around the best fit value), and the associated
parameter covariance matrix. We use this information to set a prior on these
parameters $\boldsymbol{b}$ given by a multivariate Gaussian
\begin{equation}\label{eq:mcmcprior}
p(\boldsymbol{b}) \propto \exp{\left\{-\frac{1}{2}(\boldsymbol{b} -
\boldsymbol{b}_0)^{\rm T} C^{-1} (\boldsymbol{b}-\boldsymbol{b}_0)\right\}},
\end{equation}
with a covariance matrix $C$. For the vast majority of pulsars we expect that
the uncertainties on the parameters are narrow enough that within their
ranges they all give essentially the same phase model (see discussion of phase
mismatch in \S\ref{sec:eval}). In this case the posterior on the parameters
should be dominated by this prior, therefore a good proposal distribution to
efficiently explore the space with is the same multivariate Gaussian.

\begin{figure}[!hbp]
\includegraphics[width=1.0\textwidth]{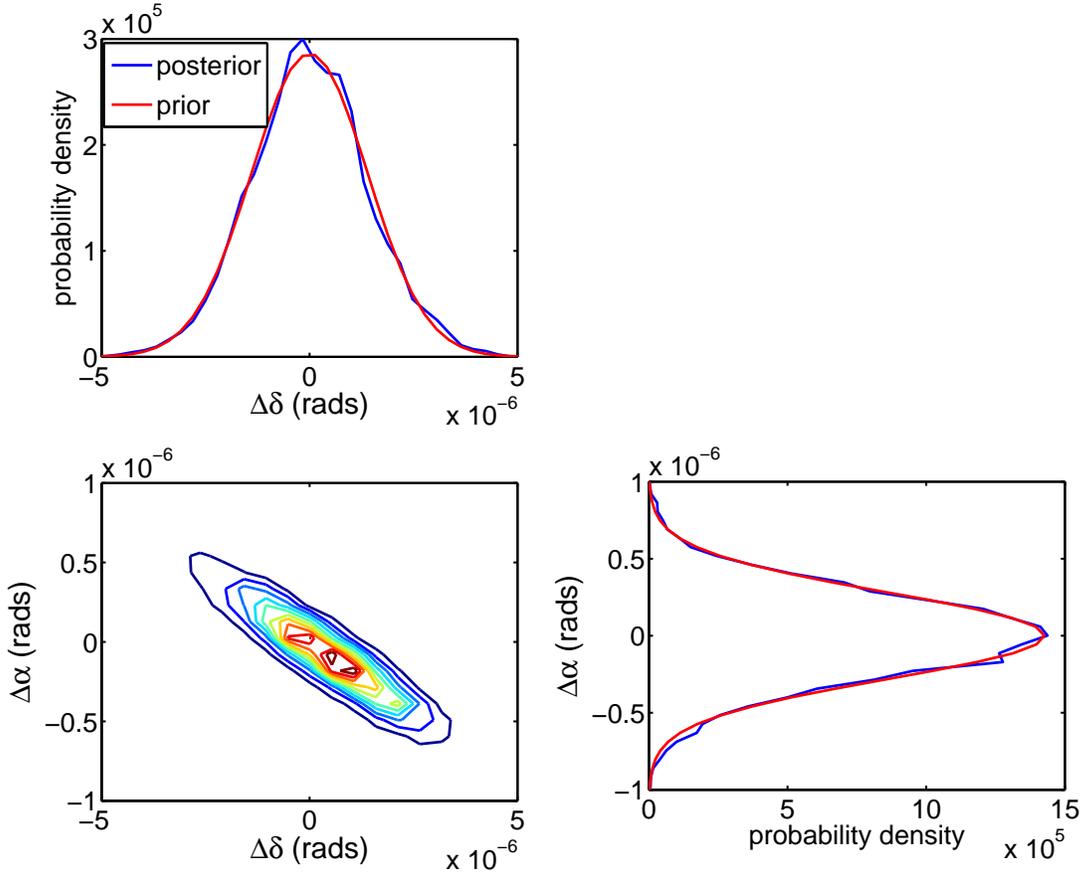}
\caption{The posteriors and priors on offsets in declination $\delta$ and right
ascension $\alpha$ computed from an MCMC search for PSR\,J0407+1607 using a day
of simulated data containing no signal. The covariance contour plot of the MCMC
chains for the two parameters is shown and has a correlation coefficient of
$-0.93$, which is identical to that of the multivariate Gaussian prior
distribution used in this study.\label{fig:correlation}}
\end{figure}

An example of the posteriors produced when searching over additional parameters
(in this case changes in declination and right ascension) can be seen in
Figure~\ref{fig:correlation}. The figure shows the multivariate Gaussian used as
priors on the two parameters and how the posterior is essentially identical to
the prior (i.e.\ the data, which contains no signal, is adding no new
information on those parameters, but the full prior space is being explored). We
have assessed this technique on many simulations of noise and found that, as
expected, the posteriors on the additional phase parameters match the priors. We
have also tested this technique on many software simulated signals (using real
pulsar phase parameters) by offsetting the injected signal parameters from the
the "best fit" values by amounts consistent with the parameter covariance
matrix. In these tests we have been able to extract the parameters as expected.

\subsection{Hardware injections}\label{sec:injections}
\begin{figure}[!hbp]
\begin{tabular}{cc}
\includegraphics[width=0.5\textwidth]{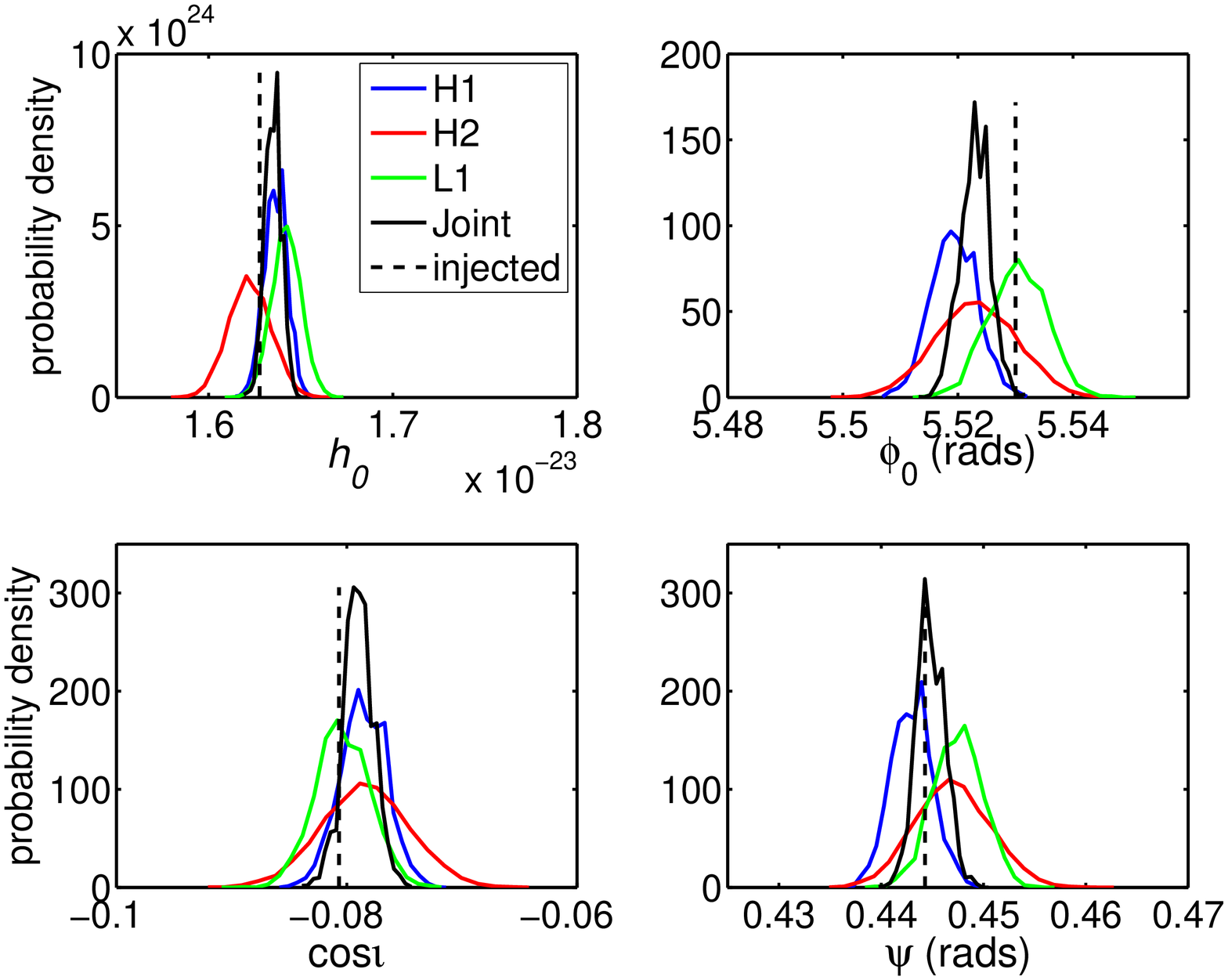} &
\includegraphics[width=0.5\textwidth]{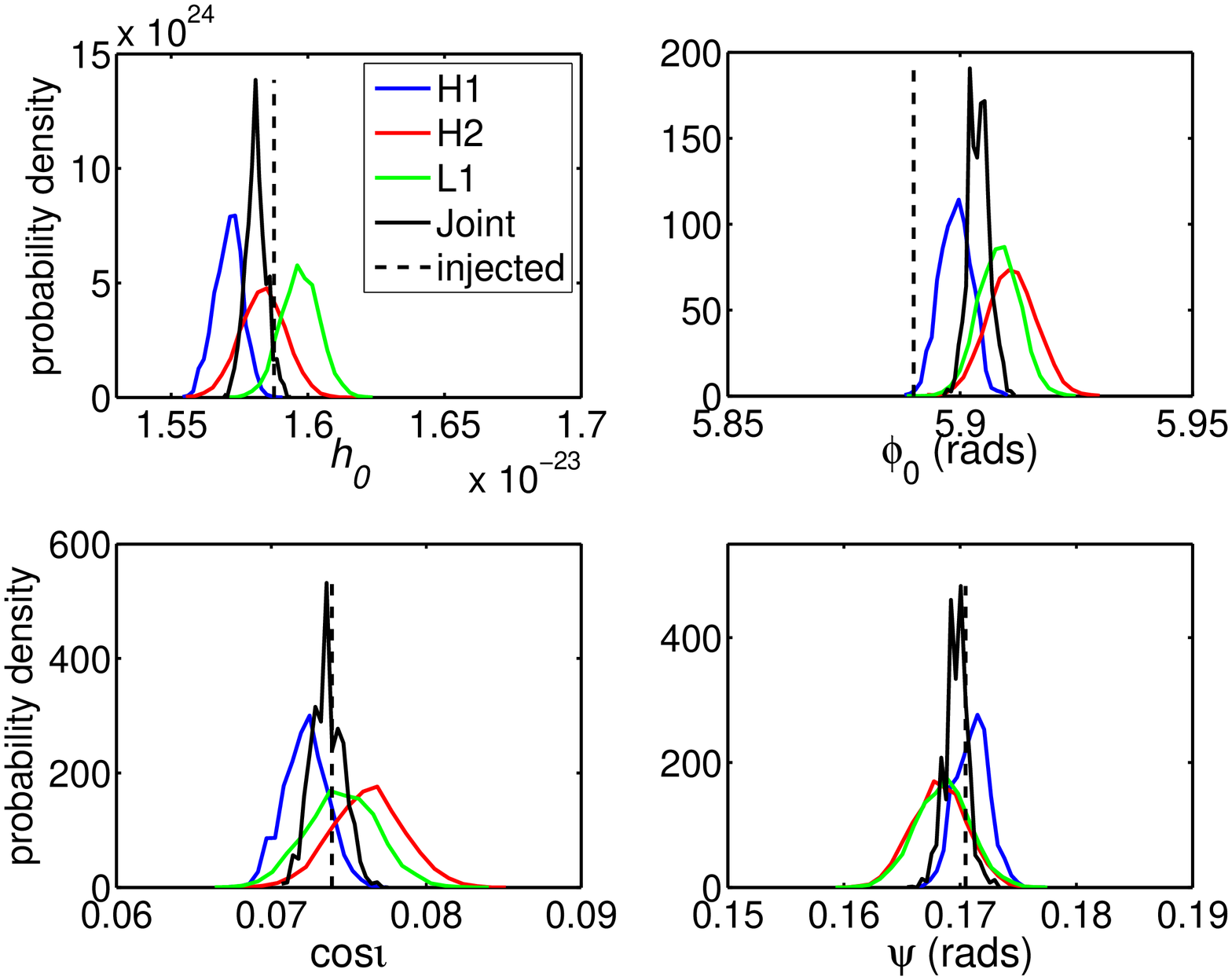}
\end{tabular}
\caption{The extracted posteriors for the pulsar parameters from the two
strongest hardware injections for each detector and for a joint detector
analysis. In each plot the injected parameter value is marked by a dashed
vertical line. The left pulsar injection was at a frequency of 108.9\,Hz
and the right pulsar was at 194.3\,Hz.\label{fig:injection}}
\end{figure}

During all LIGO science runs, except the first, fake pulsar signals have been
injected into the detectors by direct actuation of the end test masses. These
have provided end-to-end validation of the analysis codes with coherence between
the different detectors. During S5, as with S3 and S4 \citep{Abbott:2007a}, ten
signals with different source parameters were injected into the detectors. As a
demonstration of the analysis method these have been extracted using the MCMC
(over the four main parameters only) and all have been recovered with their
known injection parameter values. The extracted parameters for the two
strongest signals are shown in
Figure~\ref{fig:injection}.
From these injections it can be seen that the parameters can be extracted to a
high accuracy and consistently between detectors. The offsets in the extracted
values from the injected values are within a few percent. These are well
within the expected uncertainties of the detector calibration, which are
approximately 10\%, 10\% and 13\% in amplitude, and $4\fdg3$ (0.08 rads),
$3\fdg4$ (0.06 rads) and $2\fdg3$ (0.04 rads) in phase for H1, H2 and
L1 respectively.

\section{Evaluation of pulsar parameter errors}\label{sec:eval}
For the majority of pulsars we have parameter correlation matrices that have
been produced during the fit to the radio data, and (as discussed in
\S\ref{subsec:phaseparams}) we can use these to search over the uncertainties on
the phase parameters. For some pulsars no correlation matrix was produced with
the radio observations, so for these we instead construct conservative
correlation matrices. These assume no correlations between any parameters,
except in the case of binary systems for which we assume a correlation of 1
between the angle and time of periastron. This gives a slightly conservative
over-estimate of the parameter errors, but is still a useful approximation for
our purposes. From these correlation matrices and the given parameter standard
deviations we produce a covariance matrix for each pulsar.

Using these covariance matrices we can also assess the potential phase mismatch
that might occur between the true signal and the best fit signal used in the
heterodyne due to errors in the pulsar phase parameters (as discussed in
\S\ref{subsec:phaseparams}.) We take the mismatch (i.e.\ the loss in signal
power caused by the heterodyne phase being offset, over time, from the true
signal phase) to be
\begin{equation}\label{eq:mismatch}
M = 1 - \left(\frac{1}{T}\int_0^T
\cos{\{\phi(\boldsymbol{b}+\delta\boldsymbol{b},t) - \phi(\boldsymbol{b},
t)\}} {\rm d}t\right)^2,
\end{equation}
where $\phi$ is the phase given a vector of phase parameters $\boldsymbol{b}$
at time $t$, and $T$ is the total observation time. To get offsets in the phase
parameters, $\delta\boldsymbol{b}$, we draw random values from the multivariate
Gaussian defined by the covariance matrix. For each pulsar we drew 100\,000
random points and calculated the mean and maximum mismatch over the period of
the S5 run. The mean mismatch is a good indicator of whether the given best fit
parameter values are adequate for our search, or whether the potential offset is
large and a search including the phase parameters is entirely necessary. The
maximum mismatch represents a potential worst case in which the true signal
parameters are are at a point in parameter space approximately 4.5$\sigma$
from the best fit values (the maximum mismatch will obviously increase if one
increases the number of randomly drawn values). Of our 113 non-glitching pulsars
there are only 16 pulsars with {\it mean} mismatches greater than 1\%, three of
which are greater than 10\%: J0218+4232 at $\sim43\%$, J0024$-$7204H at
$\sim15\%$ and J1913+1011 at $\sim37\%$. Given the exclusion criterion for the
S3 and S4 analyses (see \S\ref{subsec:phaseparams}) these three pulsars would
have been removed from this analysis. There are 16 pulsars with {\it maximum}
mismatches greater than 10\% with three of these being at almost 100\%---i.e.\
if the signal parameters truly were offset from their best fit values by this
much, and these parameters were not searched over, then the signal would be
completely missed. This suggests that for the majority of pulsars the search
over the four main parameters of $h_0$, $\phi_0$, $\cos{\iota}$ and $\psi$ is
all that is necessary. But for a few, and in particular the three with large
mean mismatches (J0218+4232, J0024$-$7204H, and J1913+1011), the search over the
extra parameters is needed to ensure not losing signal power.

\section{Analysis}\label{sec:analysis}
We have used the MCMC search over {\it all} phase parameters, where they have
given errors, for all but three pulsars (see below). For the majority of pulsars
it is unnecessary (though harmless) to include these extra parameters in the
search as their priors are so narrow, but it does provide an extra demonstration
of the flexibility of the method. To double check, we also produced results for
each pulsar using the four dimensional grid of the earlier analyses and found
that, for pulsars with negligible mismatch, the results were consistent to
within a few percent. 

For our full analysis we produced three independent MCMC chains with burn-in
periods of 100\,000 iterations, followed by 100\,000 iterations to sample the
posterior. The three chains were used to assess convergence using the tests
discussed above in \S\ref{sec:mcmc}. All chains were seen to converge and were
therefore combined to give a total chain length of 300\,000 iterations from
which the posteriors were generated.

For the three pulsars that glitched during S5 (the Crab, J1952+3252 and
J0539$-$6910) the MCMC was not used to search over the position and frequency
parameters as above, as the uncertainties on these parameters gave negligable
potential mismatch. Our analysis of these pulsars did however include extra
parameters in the MCMC to take into account potential uncertainties in the
model caused by the glitches. We analysed the data coherently over the full run
as well as in stretches separated by the glitches that are treated separately.
We also included a model that allowed for a fixed but unknown phase jump
$\Delta\phi$ at the time of each glitch, keeping other physical parameters fixed
across the glitch. Results for all of these cases are given in
\S\ref{subsec:glitching}.

The maximum detector calibration uncertainties are approximately 10\%, 10\% and
13\% in amplitude, and $4\fdg3$ (0.08 rads), $3\fdg4$ (0.06 rads) and $2\fdg3$
(0.04 rads) in phase for H1, H2 and L1 respectively.

\section{Results}\label{sec:results}
No evidence of a \gw signal was seen for any of the pulsars. In light of this,
we present joint 95\% upper limits on $h_0$ for each pulsar (see
Table~\ref{tab:results} for the results for the non-glitching pulsars). We also
interpret these as limits on the pulsar ellipticity, given by
\begin{equation}\label{eq:ellipticity}
\varepsilon = 0.237\,h_{-24}r_{\rm kpc}\nu^{-2}I_{38},
\end{equation}
where $h_{-24}$ is the $h_0$ upper limit in units of $1\ee{-24}$, $I_{38}=1$ and
$r_{\rm kpc}$ pulsar distance in kpc. For the majority of pulsars this distance
is taken as the estimate value from Australia Telescope National Facility Pulsar
Catalogue\footnote{\url{http://www.atnf.csiro.au/research/pulsar/psrcat/}}
\citep{Manchester:2005}, but for others more up-to-date distances are known. For
pulsars in Terzan 5 (those with the name J1748$-$2446) a distance of 5.5\,kpc is
used \citep{Ortolani:2007}. For J1939+2134 the best estimate distance is a
highly uncertain value of $\sim8.3\pm5$\,kpc based on parallax measurements
\citep{Kaspi:1994}, but it is thought to be a large overestimate, so instead we
use a value of 3.55\,kpc derived from the \citet{Cordes:2002} NE2001 galactic
electron density model. The observed spin-down rate for globular cluster pulsars
is contaminated by the accelerations within the cluster, which can lead to some
seeming to spin-up as can be seen in column four of Table~\ref{tab:results}.
For all pulsars in globular clusters, except J1824$-$2452A, we instead calculate
a conservative spin-down limit by assuming all pulsars have a characteristic age
$\tau = \nu/2\dot{\nu}$ of $10^9$\,years. Note that using the characteristic age
gives a spin-down limit that is independent of frequency and only depends on
$\tau$ and $r$. Globular cluster pulsar J1824$-$2452A has a large spin-down 
that is well above what could be masked by cluster accelerations. Therefore, for
this pulsar we use its true spin-down for our limit calculation. For some nearby
pulsars there is a small, but measurable, Shklovskii effect which will 
contaminate the observed spin-down. For these if a value of the intrinsic 
spin-down is known then this is used when calculating the spin-down limit.

\begin{figure}[!hbp]
\includegraphics[width=1.0\textwidth]{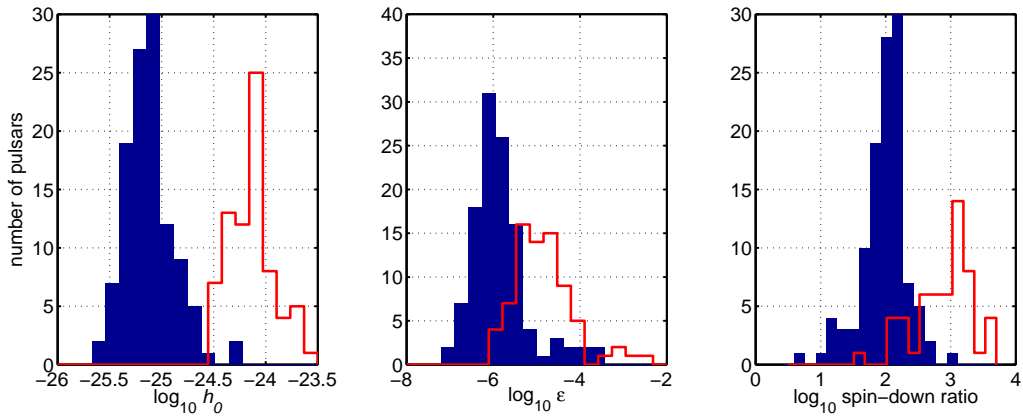}
\caption{The solid histograms show the results of this analysis in terms of
upper limits on $h_0$, the ellipticity $\varepsilon$ and the ratio to the
spin-down limit (excluding the glitching pulsars). The clear histograms show the
same set of values for the combined S3 and S4 analysis
\citep{Abbott:2007a}.\label{fig:histograms}}
\end{figure}

\begin{figure}[!hbp]
\includegraphics[width=1.0\textwidth]{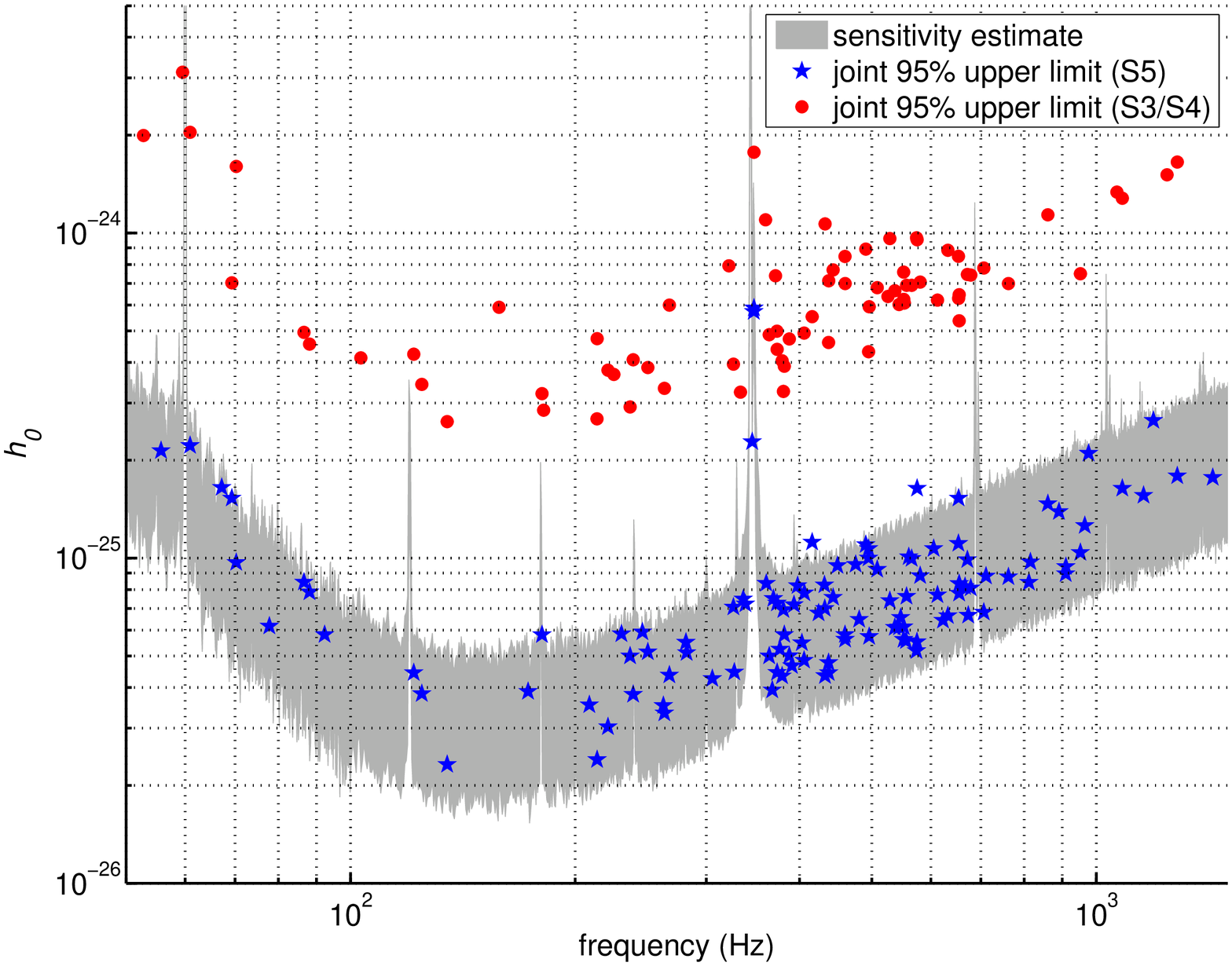}
\caption{The \gw amplitude upper limits are plotted over the estimated
sensitivity of the search as defined by the grey band (see text). Also plotted 
are the limits from the S3/S4 search.\label{fig:sensestimate}}
\end{figure}

The results are plotted in histogram form in Figure~\ref{fig:histograms}. Also
shown for comparison are the results of the previous search using combined data
from the S3 and S4 science runs. The median upper limit on $h_0$ for this
search, at $7.2\ee{-26}$, is about an order of magnitude better than in the
previous analysis ($6.3\ee{-25}$). A large part of this increased sensitivity
(about a factor of 4 to 5) is due the the longer observation time. The median
ellipticity is $\varepsilon = 1.1\ee{-6}$, an improvement from $9.1\ee{-6}$, and
the median ratio to the spin-down limit is 108, improved from 870. If one
excludes the conservatively estimated spin-down limits for the globular cluster
pulsars, the median ratio is 73. In Figure~\ref{fig:sensestimate} the upper
limits on $h_0$ are also plotted overlaid onto an estimate of the search
sensitivity\footnote{The upper and lower estimated sensitivity limits for the
shaded band in Figure~\ref{fig:sensestimate} come from the values of the 95\%
$h_0$ upper limits that bound 95\% of total values from simulations on white
noise for randomly distributed pulsars. The band can be estimated from Figure 1
of \citet{Dupuis:2005}, which gives limits of (7 to 20)$\times \sqrt{S_n/T}$,
where $S_n$ is the single sided power spectral density and $T$ is the total
observation time in seconds.}. The expected uncertainties in these results due
to the calibrations are given in \S\ref{sec:injections}.

The smallest upper limit on $h_0$ for any pulsar is $2.3\ee{-26}$ for
J1603$-$7202, which has a \gw frequency of 135\,Hz and is in the most sensitive
part of the detectors' bands. The lowest ellipticity upper limit is $7.0\ee{-8}$
for J2124$-$3358, which has a \gw frequency of 406\,Hz and a best estimate
distance of 0.2\,kpc. Of the millisecond recycled pulsars this is also the
closest to its spin-down limit, at a value of 9.4 times greater than this
limit. Of all pulsars which did not glitch during S5, the young pulsar
J1913+1011 is the closest to its spin-down limit, at only 3.9 times greater than
it.

\begin{deluxetable}{ c  c  c  c  c  c  c  c  c }
\rotate
\tablewidth{0pt}
\tabletypesize{\footnotesize}
\tablecaption{Information on the non-glitching pulsars in our search,
including the start and end times of the radio observations used in producing
the parameter fits for our search, and upper limit results.\label{tab:results}}
\tablehead{\colhead{Pulsar} & \colhead{start -- end (MJD)} & \colhead{$\nu$
(Hz)} & \colhead{$\dot{\nu}$ (Hz\,s$^{-1}$)} & \colhead{distance (kpc)} &
\colhead{spin-down limit} & \colhead{joint $h_0^{95\%}$} &
\colhead{ellipticity} & \colhead{$h_0^{95\%}/h_0^{\rm sd}$}}
\startdata
\input pulsartable.tex
\enddata
\tablenotetext{b}{The pulsar is within a binary system.}
\tablenotetext{c}{The pulsar is within a globular cluster.}
\tablenotetext{g}{The pulsar was observed by the Green Bank Telescope.}
\tablenotetext{j}{The pulsar was observed by the Jodrell Bank Observatory.}
\tablenotetext{p}{The pulsar was observed by the Parkes Observatory.}
\tablenotetext{\dagger}{The pulsar's spin-down is corrected for proper motion
effects}
\end{deluxetable}

\subsection{Glitching pulsars}\label{subsec:glitching}
For the three pulsars that glitched, we have chosen to take into account three
different models related to the coherence of the \gw signal and the
electromagnetic signal over the glitch: i) there is coherence between them over
the glitch (i.e.\ the glitch causes no discontinuity between the electromagnetic
and \gw phases); ii) there is decoherence (in terms of a phase jump) between
them at the time of the glitch, but the phase discontinuity is included as an
extra search parameter (i.e.\ a $\Delta\phi$ parameter is added at the time of
the glitch); and iii) the data stretches before, between and after the glitches
are treated separately and analysed independently.

\subsubsection{Crab pulsar}\label{subsec:crab}
For this search we include the three different models described above
relating to the observed glitch in the pulsar on 2006 August 3. Potentially all
the four main \gw signal parameters could be changed during the glitch if it is
large enough to cause major disruption to the star, but this is not the case for
the observed glitch. It had a fractional frequency change of order
$\Delta{}\nu/\nu \sim 5\ee{-9}$, which is unlikely to be energetic enough to
cause changes in the \gw amplitude near our current levels of sensitivity. To
model the signal phase we use the regularly updated Crab pulsar Monthly
Ephemeris \citep{CrabEphemerisPaper, CrabEphemeris}, which is needed to take
into account the phase variations caused by timing noise.

As for the previous Crab Pulsar search \citep{Abbott:2008} we have information
on the orientation of the pulsar from the orientation of the pulsar wind nebula
(PWN) \citep{Ng:2008}. We use this information to set Gaussian priors on the
$\psi$ and $\iota$ parameters of $\psi = 125\fdg155\pm1\fdg355$ (the
$\psi$-dependence wraps around at $\pm45^{\circ}$, so the actual value used is
$35\fdg155$) and $\iota=62\fdg17\pm2\fdg20$.

There is reason to believe that the PWN orientation reflects that of the central
pulsar, but in case this is not an accurate description we present
results using both a uniform prior over all parameters, and
using the restricted prior ranges. The MCMC only searches over the four main
parameters and does not include errors on the pulsar position, frequency or
frequency derivatives as these are negligible. The results are summarised in
Table~\ref{tab:crabresults}, where for the ellipticity and spin-down limit
calculations a distance of 2\,kpc was used. The orientation angle suggested by
the restricted priors is slightly favourable in terms of the observable
gravitational wave emission it would produce, and this allows us to set a better
upper limit.

In Figure~\ref{fig:momellplane} we plot the result from model i), with
restricted priors, as an exclusion region on the moment of inertia--ellipticity
plane. It can been seen that for all the allowed regions in moment of inertia
we beat the spin-down limit. If one assumes the Crab pulsar to have a moment of
inertia at the upper end of the allowed range our result beats the spin-down
limit by over an order of magnitude.

\subsubsection{PSR\,J0537$-$6910}
For pulsar J0537$-$6910, which has so far been timed only in X-rays, we rely
on data from the Rossi X-ray Timing Explorer (RXTE) satellite. This pulsar is
young, has a high spin-down rate, and is a prolific glitcher, and therefore we
require observations overlapping with our data to produce a coherent
template. \citet{Middleditch:2006} have published observations covering from the
beginning of S5 up to 2006 August 21, during which time the pulsar was seen to
glitch three times. Further observations have been made which span the rest of
the S5 run and show another three glitches during this time. The epochs and
parameters for the seven ephemeris periods overlapping with our data run are
given in Table~\ref{tab:0537epochs}. For the first epoch there was no data for 
L1, so the joint result only uses H1 and H2 data. For the analyses using all the 
data (models i and ii) we have 474 days of H1 data, 475 days of H2 data and 397
days of L1 data. Due to the glitches we perform parameter estimation for the
same three models given above.
\begin{deluxetable}{c c c c c c}
\tablecaption{Ephemeris information for PSR\,J0537$-$6910. The first four values
are taken from \citet{Middleditch:2006}.\label{tab:0537epochs}}
\tabletypesize{\footnotesize}
\tablewidth{0pt}
\tablehead{\colhead{~} & \colhead{start -- end (MJD)} & \colhead{$\nu$ (Hz)} &
\colhead{$\dot{\nu}$ (Hz\,s$^{-1}$)} & \colhead{$\ddot{\nu}$ (Hz\,s$^{-2}$)} &
\colhead{epoch (MJD)}}
\tablewidth{0pt}
\startdata
1. & 53551--53687 & 62.000663106 & $-1.994517\ee{-10}$ & $11.2\ee{-21}$ &
53557.044381976239 \\
2. & 53711--53859 & 61.996292178 & $-1.993782\ee{-10}$ & $8.1\ee{-21}$ &
53812.224185839386 \\
3. & 53862--53950 & 61.995120229 & $-1.994544\ee{-10}$ & $9.6\ee{-21}$ &
53881.096123033291 \\
4. & 53953--53996 & 61.993888026 & $-1.995140\ee{-10}$ & $9.6\ee{-21}$ &
53952.687378384607 \\
5. & 54003--54088 & 61.992869785 & $-1.994516\ee{-10}$ & $2.2\ee{-21}$ &
54013.061576594146 \\
6. & 54116--54273 & 61.990885506 & $-1.994577\ee{-10}$ & $8.1\ee{-21}$ &
54129.540333159754 \\
7. & 54277--54441 & 61.988307647 & $-1.994958\ee{-10}$ & $7.7\ee{-21}$ &
54280.918705402011 \\
\enddata
\end{deluxetable}

As with the Crab Pulsar there is also information on the orientation of
J0537$-$6910 from model fits to its pulsar wind nebula \citep{Ng:2008}. These
are used to set Gaussian priors on $\psi$ and $\iota$ of
$\psi=131\fdg0\pm2\fdg2$ (equivalently $41\fdg0$) and $\iota=92\fdg8\pm0\fdg9$.
We again quote results using uniform priors over all parameters and with these
restricted priors. The distance used in the ellipticity and spin-down limits for
J0537$-$6910 is 49.4\,kpc. The results are summarised in
Table~\ref{tab:crabresults}. Using the restricted priors we obtain a worse upper
limit on $h_0$ than for uniform priors. This is because the nebula suggests that
the star has its spin axis perpendicular to the line of sight, and therefore the
gravitational radiation is linearly polarised and the numerical strain amplitude
is lower than average for a given strain tensor amplitude $h_0$.

In Figure~\ref{fig:momellplane} we again plot the result from model i), with
restricted priors, on the moment of inertia--ellipticity plane. It can be seen
that the spin-down limit is beaten if we assume a moment of inertia to be 
$2\times10^{38}$\,kg\,m$^2$ or greater.

\subsubsection{PSR\,J1952+3252}\label{subsec:J1952}
\begin{figure}[!hbp]
\includegraphics[width=0.7\textwidth]{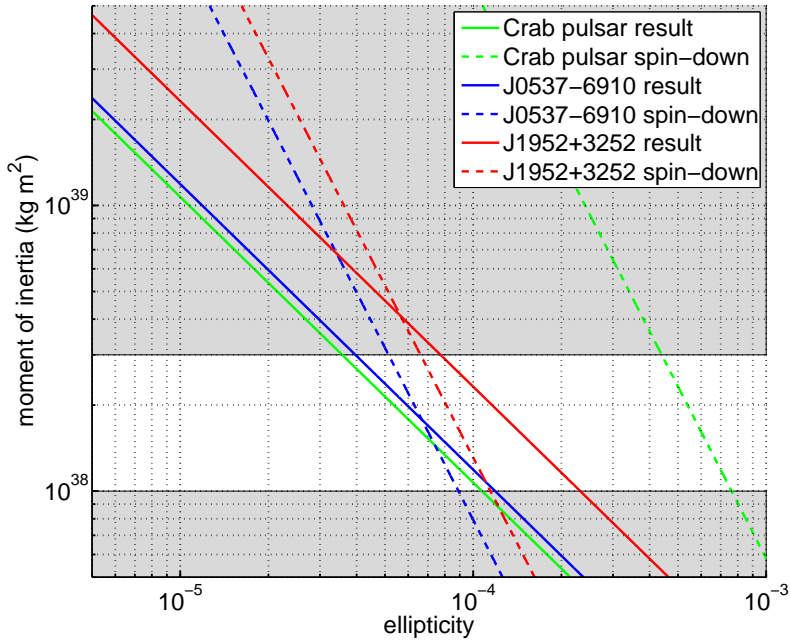}
\caption{The results of the Crab pulsar, J0537$-$6910 and J1952+3252 analyses,
and the spin-down limits, plotted on the moment of inertia--ellipticity plane.
The results used are those from model i) and with restricted priors on the angular
parameters for the Crab pulsar and J0537$-$6910. Areas to the right of the
diagonal lines are excluded. The shaded regions are those outside the
theoretically predicted range of moments of inertia $I_{38} =
$1--3.\label{fig:momellplane}}
\end{figure}

PSR\,J1952+3252 is another young pulsar with a high spin-down rate (although a
couple of orders of magnitude less than for the Crab pulsar and J0537$-$6910) --
it has a spin parameters of $\nu = 25.30$\,Hz and
$\dot{\nu}=-3.73\ee{-12}$\,Hz\,s$^{-1}$. Jodrell Bank observations of this
pulsar were made over the whole of S5, but it was observed to glitch at some
point between 2007 January 1 and January 12. For both the pre and post-glitch
epochs we have coherent timing solutions and again perform analyses as above.
The are no constraints on the orientation of this pulsar, so we do not use any
restricted priors. The results for this pulsar are given in
Table~\ref{tab:crabresults}. We reach about a factor of two above the
spin-down limit using a distance of 2.5\,kpc. The result from model i) is
also plotted on the moment of inertia--ellipticity plane in
Figure~\ref{fig:momellplane}. It can be seen that $I_{38}$ just over
over 4, which is above the expected maximum allowable value,
would be needed to beat the spin-down limit.

\begin{deluxetable}{c c c c c c c}
\tablecaption{Results of the analysis for the Crab pulsar, J0537$-$6910 and
J1952+3252.\label{tab:crabresults}}
\tabletypesize{\footnotesize}
\tablewidth{0pt}
\tablehead{\colhead{~} & \multicolumn{2}{c}{$h_0^{95\%}$} &
\multicolumn{2}{c}{ellipticity} & \multicolumn{2}{c}{$h_0^{95\%}/h_0^{\rm sd}$}
\\
\colhead{epoch} & \colhead{uniform} &
\colhead{restricted\tablenotemark{\dagger}} & \colhead{uniform} &
\colhead{restricted\tablenotemark{\dagger}} & \colhead{uniform} &
\colhead{restricted\tablenotemark{\dagger}}}
\tablecolumns{7}
\startdata
\cutinhead{Crab pulsar}
model i)\tablenotemark{a} & $2.6\ee{-25}$ & $2.0\ee{-25}$ & $1.4\ee{-4}$ &
$1.1\ee{-4}$ & 0.18 & 0.14 \\
model ii)\tablenotemark{b} & $2.4\ee{-25}$ & $1.9\ee{-25}$ & $1.3\ee{-4}$ &
$9.9\ee{-5}$ & 0.17 & 0.13 \\
1. & $4.9\ee{-25}$ & $3.9\ee{-25}$ & $2.6\ee{-4}$ & $2.1\ee{-4}$ & 0.34 & 0.27
\\
2. & $2.4\ee{-25}$ & $1.9\ee{-25}$ & $1.3\ee{-4}$ & $1.0\ee{-4}$ & 0.15 & 0.13
\\
\cutinhead{J0537$-$6910}
model i)\tablenotemark{a} & $2.9\ee{-26}$ & $3.9\ee{-26}$ & $8.9\ee{-5}$ &
$1.2\ee{-4}$ & 1.0 & 1.3 \\
model ii)\tablenotemark{b} & $4.1\ee{-26}$ & $4.6\ee{-26}$ & $1.2\ee{-4}$ &
$1.4\ee{-4}$ & 1.4 & 1.5 \\
1. & $9.1\ee{-25}$ & $1.2\ee{-24}$ & $2.8\ee{-3}$ & $3.7\ee{-3}$ & 31.2 & 41.2
\\
2. & $1.2\ee{-25}$ & $1.5\ee{-25}$ & $3.6\ee{-4}$ & $4.6\ee{-4}$ & 4.1 & 5.2
\\
3. & $1.4\ee{-25}$ & $1.3\ee{-25}$ & $4.2\ee{-4}$ & $3.8\ee{-4}$ & 4.7 & 4.3
\\
4. & $8.3\ee{-26}$ & $1.1\ee{-25}$ & $2.5\ee{-4}$ & $3.3\ee{-4}$ & 2.8 & 3.7
\\
5. & $1.4\ee{-25}$ & $1.3\ee{-25}$ & $4.1\ee{-4}$ & $4.0\ee{-4}$ & 4.7 & 4.6
\\
6. & $4.4\ee{-26}$ & $5.8\ee{-26}$ & $1.3\ee{-4}$ & $1.8\ee{-4}$ & 1.5 & 2.0
\\
7. & $4.9\ee{-26}$ & $6.1\ee{-26}$ & $1.5\ee{-4}$ & $1.9\ee{-4}$ & 1.7 & 2.1
\\
\cutinhead{J1952+3252}
model i)\tablenotemark{a} & $2.5\ee{-25}$ & \nodata & $2.3\ee{-4}$ & \nodata
& 2.0 & \nodata \\
model ii)\tablenotemark{b} & $3.6\ee{-25}$ & \nodata & $3.3\ee{-4}$ & \nodata
& 2.9 & \nodata \\
1. & $4.7\ee{-25}$ & \nodata & $4.3\ee{-4}$ & \nodata & 3.8 & \nodata \\
2. & $4.4\ee{-25}$ & \nodata & $4.0\ee{-4}$ & \nodata & 3.5 & \nodata \\
\enddata
\tablenotetext{\dagger}{Uses observationally restricted priors on the
orientation angle and polarisation angle}
\tablenotetext{a}{Uses the full data set for a coherent analysis}
\tablenotetext{b}{Uses the full data set, but searches over an extra phase
parameter at each glitch}
\end{deluxetable}

\section{Conclusions}
In this paper we have searched for continuous gravitational waves from an
unprecedented number of pulsars with unprecedented sensitivity, using coincident
electromagnetic observations of many millisecond and young pulsars. Our direct
upper limits have beaten the indirect limits (including the spin-down limit)
for the Crab Pulsar and are at the canonical spin-down limit for J0537$-$6910.
For several more pulsars our upper limits are within roughly one order of
magnitude of the spin-down limits, and therefore we expect a comparable search
of Advanced LIGO and Virgo data to beat the latter.

For the Crab Pulsar we improve upon the previous upper limit, which already beat
the spin-down limit and other indirect limits; and we also provide results for
different scenarios in which the signals may not be fully coherent over the run.
By assuming the observationally constrained priors on the angular parameters are
correct we can limit the power output of the pulsar (see \citet{Abbott:2008}) to
be less than $\sim2\%$ of the canonical spin-down luminosity.

Pulsar J0537$-$6910 is an interesting case due to its youth and glitchiness.
Assuming that the \gw signal phase is coherent with the X-ray pulses over
glitches, our upper limit is within a few percent of the canonical spin-down
limit for this pulsar. If we assume that the phase is incoherent over the
glitches then using the full data set our upper limit is $\sim1.2$ times the
canonical spin-down limit. Even if we assume that none of the four main
pulsar parameters remains coherent during a glitch (i.e.\ there is a major
rearrangement in the star's structure) then the results from the final two 
epochs individually are still only approximately two times the spin-down limit. 
All of these direct upper limits are within the uncertainties of the spin-down 
limit, and (unlike the Crab) there is no firmly observed braking index and thus 
no substantially stricter indirect limit.

Pulsar J1952+3252 gets close to its spin-down limit, but even taking
into account moment of inertia and distance uncertainties, probably would still
just miss beating it. It does however make an exciting target for future runs.

Our observational upper limits on the ellipticities of the Crab and J0537$-$6910
are in the vicinity of $10^{-4}$. Elastic deformations of this magnitude have
been predicted as sustainable \citep{Owen:2005, Haskell:2007zz, Lin:2007rz,
Knippel:2009st} not for normal neutron stars but only for exotic forms of
crystalline quark matter \citep{Xu:2003xe, Mannarelli:2007bs}. Such
ellipticities could also be sustained by internal toroidal magnetic fields of
order $10^{16}$~G depending on the field configuration, equation of state, and
superconductivity of the star \citep{Cutler:2002, Akgun:2007ph, Haskell:2007bh,
Colaiuda:2007br}. Therefore our observations have achieved the sensitivity to
detect some of the more extreme possibilities, and have constrained the internal
magnetic fields of the Crab and J0537$-$6910 to be less than of order
$10^{16}$~G. However, it is important to be clear that the upper limits reported
here do not constrain the properties of crystalline quark matter, because
gravitational wave observations constrain the true ellipticity rather than the
maximum ellipticity.  This distinction was made clear in \citet{Owen:2005}, but
has not been properly enforced in more recent work (\citep{Lin:2007rz,
Knippel:2009st}.  Also, while it is the case that, as anticipated by
\citet{Haskell:2007zz}, the upper bounds on ellipticity reported here for some
pulsars, notably J2124-3358, have pushed into a regime where elastic strains
might support such deformations, detection was not expected, as such large
ellipticities would conflict with the spin-down limit.

Recent work by \cite{Horowitz:2009} suggests that the breaking strain of
neutron star crusts may be an order of magnitude higher ($10^{-1}$) than the
highest values ($10^{-2}$) previously used in estimates of maximum sustainable
ellipticities. Their simulations are strictly applicable only to the outer crust
(i.e.\ no neutron drip), but since the reason for the high breaking strain is
very generic---the extreme pressure simply crushes away defects that contribute
to early fracture---it may apply to the inner crust (the major contributor
to ellipticity) as well. In that case normal neutron stars could sustain
ellipticities close to $10^{-5}$, which is beyond reach of the present search
(for those stars where our results are at or near the spin-down limit)
but would be accessible with data from advanced interferometers. If the high
breaking strain holds for mixed phases in hybrid stars (which is not clear due
to the increasing importance of the strong interaction at high densities), it
would bring their maximum ellipticities estimated by \citet{Owen:2005} up an
order of magnitude to about the $10^{-4}$ values achieved here.

In 2009 July the LIGO 4\,km detectors (featuring some upgraded systems
and titled Enhanced LIGO) and the upgraded Virgo detector began their sixth
and second science runs respectively (S6 and VSR2). These upgrades aim to
provide significantly better strain sensitivities than S5/VSR1. For LIGO these
improvements will primarily be at frequencies greater than 40\,Hz, but Virgo
will improve at lower frequencies too and should outperform LIGO below about
40\,Hz. LIGO and Virgo will closely approach and could potentially beat the
spin-down limits not only for the Crab and J0537-6910 but for six more known
pulsars: J1952+3252, J1913+1011, J0737$-$3039A, J0437$-$4715\footnote{This
pulsar sits on a strong LIGO spectral noise line caused by the violin mode
resonance of the suspension wires, so the required sensitivity may only be
reachable with Virgo.}, and the recently discovered J1747$-$2809
\citep{Camilo:2009gd} and J1813$-$1749 \citep{Gotthelf:2009}. Below 40\,Hz Virgo
could beat the spin-down limits for three more pulsars: the Vela pulsar,
J0205+6449 and J1833$-$1034. Most of these eleven pulsars are young and may be
prone to large levels of timing noise and glitches. Therefore to reach the best
sensitivities it is essential to have radio and X-ray observations of these
sources made in coincidence with the S6/VSR2 run. In particular the extension of
the RXTE satellite is crucial to our ability to perform targeted searches of
J0537$-$6910 during this time of enhanced detector performance.

On a time scale of a few years, the Advanced LIGO and Virgo interferometers are
expected to achieve strain sensitivities more than 10 times better for the
pulsars in the current band, and will extend that band downward to approximately
10\,Hz. Searches at such sensitivities will beat the spin-down limits from
dozens of known pulsars and also enter the range of ellipticities predicted for
normal neutron stars, improving the prospects for direct \gw observations from
these objects.

\acknowledgements
The authors gratefully acknowledge the support of the United States National
Science Foundation for the construction and operation of the LIGO Laboratory,
the Science and Technology Facilities Council of the United Kingdom, the
Max-Planck-Society, and the State of Niedersachsen/Germany for support of the
construction and operation of the GEO600 detector, and the Italian Istituto
Nazionale di Fisica Nucleare and the French Centre National de la Recherche
Scientifique for the construction and operation of the Virgo detector. The
authors also gratefully acknowledge the support of the research by these 
agencies and by the Australian Research Council, the Council of Scientific and
Industrial Research of India, the Istituto Nazionale di Fisica Nucleare of
Italy, the Spanish Ministerio de Educaci\'on y Ciencia, the Conselleria
d'Economia Hisenda i Innovaci\'o of the Govern de les Illes Balears, the
Foundation for Fundamental Research on Matter supported by the Netherlands
Organisation for Scientific Research, the Royal Society, the Scottish Funding
Council, the Polish Ministry of Science and Higher Education, the FOCUS
Programme of Foundation for Polish Science, the Scottish Universities Physics
Alliance, The National Aeronautics and Space Administration, the Carnegie Trust,
the Leverhulme Trust, the David and Lucile Packard Foundation, the Research
Corporation, and the Alfred P. Sloan Foundation. LIGO Document No.
LIGO-P080112-v5.

Pulsar research at UBC is supported by a Natural Sciences and Engineering
Research Council of Canada Discovery Grant. The Parkes radio telescope is part
of the Australia Telescope which is funded by the Commonwealth Government for
operation as a National Facility managed by CSIRO. The National Radio Astronomy 
Observatory is a facility of the United States National Science Foundation 
operated under cooperative agreement by Associated Universities, Inc. We thank
Maura McLaughlin for useful discussions.
\bibliography{S5KnownPulsarPaper}

\end{document}

%% file: authorlist.tex

\author{B.~P.~Abbott\altaffilmark{28}, 
R.~Abbott\altaffilmark{28}, 
F.~Acernese\altaffilmark{18ac}, 
R.~Adhikari\altaffilmark{28}, 
P.~Ajith\altaffilmark{2}, 
B.~Allen\altaffilmark{2,75}, 
G.~Allen\altaffilmark{51}, 
M.~Alshourbagy\altaffilmark{20ab}, 
R.~S.~Amin\altaffilmark{33}, 
S.~B.~Anderson\altaffilmark{28}
% ...
}

\affil{The LIGO Scientific Collaboration \& The Virgo Collaboration}

% ...
\altaffiltext{2}{Albert-Einstein-Institut, Max-Planck-Institut f\"ur Gravitationsphysik, D-30167 Hannover, Germany}
% ...
\altaffiltext{18}{INFN, sezione di Napoli $^a$; Universit\`a di Napoli 'Federico II'$^b$ Complesso Universitario di Monte S.Angelo, I-80126 Napoli; Universit\`a di Salerno, Fisciano, I-84084 Salerno$^c$, Italy}
% ...
\altaffiltext{20}{INFN, Sezione di Pisa$^a$; Universit\`a di Pisa$^b$; I-56127 Pisa; Universit\`a di Siena, I-53100 Siena$^c$, Italy}
% ...
\altaffiltext{28}{LIGO - California Institute of Technology, Pasadena, CA  91125, USA }
% ...
\altaffiltext{33}{Louisiana State University, Baton Rouge, LA  70803, USA }
% ...
\altaffiltext{51}{Stanford University, Stanford, CA  94305, USA }
% ...
\altaffiltext{75}{University of Wisconsin-Milwaukee, Milwaukee, WI  53201, USA }
% ...

\author{S.~B\'egin\altaffilmark{86, 89},
A.~Corongiu\altaffilmark{82},
N.~D'Amico\altaffilmark{82 ,81},
P.~C.~C.~Freire\altaffilmark{78, 90},
J.~W.~T.~Hessels\altaffilmark{85, 79},
G.~B.~Hobbs\altaffilmark{80},
M.~Kramer\altaffilmark{87},
A.~G.~Lyne\altaffilmark{87}, 
R.~N.~Manchester\altaffilmark{80},
F.~E.~Marshall\altaffilmark{88},
J.~Middleditch\altaffilmark{83},
A.~Possenti\altaffilmark{82},
S.~M.~Ransom\altaffilmark{84},
I.~H.~Stairs\altaffilmark{86},
and B.~Stappers\altaffilmark{87}}

\altaffiltext{78}{Arecibo Observatory, HC 3 Box 53995, Arecibo, Puerto Rico
00612, USA}
\altaffiltext{79}{Astronomical Institute ``Anton Pannekoek'', University 
of Amsterdam, 1098 SJ Amsterdam, The Netherlands}
\altaffiltext{80}{Australia Telescope National Facility, CSIRO, PO Box 76,
Epping NSW 1710, Australia}
\altaffiltext{81}{Dipartimento di Fisica Universit\`a di Cagliari,
Cittadella Universitaria, I-09042 Monserrato, Italy}
\altaffiltext{82}{INAF - Osservatorio Astronomico di Cagliari, Poggio dei Pini,
09012 Capoterra, Italy}
\altaffiltext{83}{Modeling, Algorithms, and Informatics, CCS-3, MS B265,
Computer, Computational, and Statistical Sciences Division, Los Alamos National
Laboratory, Los Alamos, NM 87545, USA}
\altaffiltext{84}{National Radio Astronomy Observatory, Charlottesville, VA
22903, USA}
\altaffiltext{85}{Netherlands Institute for Radio Astronomy (ASTRON), Postbus 2, 
7990 AA Dwingeloo, The Netherlands}
\altaffiltext{86}{Department of Physics and Astronomy, University of British
Columbia, 6224 Agricultural Road, Vancouver, BC V6T 1Z1, Canada}
\altaffiltext{87}{University of Manchester, Jodrell Bank Centre for Astrophysics
Alan-Turing Building, Oxford Road, Manchester M13 9PL, UK}
\altaffiltext{88}{NASA Goddard Space Flight Center, Greenbelt, MD 20771, USA}
\altaffiltext{89}{D\'epartement de physique, de g\'enie physique et d'optique, 
Universit\'e Laval, Qu\'ebec, QC G1K 7P4, Canada.}
\altaffiltext{90}{West Virginia University, Department of Physics, PO Box 6315,
Morgantown, WV 26506, USA}

%% file: pulsartable.tex
J0024$-$7204C\tablenotemark{cp} & 48383 -- 54261 & 173.71 & $1.5\times10^{-15}$ &4.9 & $6.55\times10^{-28}$ &$5.88\times10^{-25}$ &$2.26\times10^{-5}$ &898 \\
J0024$-$7204D\tablenotemark{cp} & 48465 -- 54261 & 186.65 & $1.2\times10^{-16}$ &4.9 & $6.55\times10^{-28}$ &$4.45\times10^{-26}$ &$1.48\times10^{-6}$ &68 \\
J0024$-$7204E\tablenotemark{bcp} & 48465 -- 54261 & 282.78 & $-7.9\times10^{-15}$ &4.9 & $6.55\times10^{-28}$ &$9.97\times10^{-26}$ &$1.44\times10^{-6}$ &152 \\
J0024$-$7204F\tablenotemark{cp} & 48465 -- 54261 & 381.16 & $-9.4\times10^{-15}$ &4.9 & $6.55\times10^{-28}$ &$8.76\times10^{-26}$ &$6.98\times10^{-7}$ &134 \\
J0024$-$7204G\tablenotemark{cp} & 48600 -- 54261 & 247.50 & $2.6\times10^{-15}$ &4.9 & $6.55\times10^{-28}$ &$1.00\times10^{-25}$ &$1.90\times10^{-6}$ &153 \\
J0024$-$7204H\tablenotemark{bcp} & 48518 -- 54261 & 311.49 & $1.8\times10^{-16}$ &4.9 & $6.55\times10^{-28}$ &$6.44\times10^{-26}$ &$7.69\times10^{-7}$ &98 \\
J0024$-$7204I\tablenotemark{bcp} & 50684 -- 54261 & 286.94 & $3.8\times10^{-15}$ &4.9 & $6.55\times10^{-28}$ &$5.19\times10^{-26}$ &$7.30\times10^{-7}$ &79 \\
J0024$-$7204J\tablenotemark{bcp} & 48383 -- 54261 & 476.05 & $2.2\times10^{-15}$ &4.9 & $6.55\times10^{-28}$ &$1.04\times10^{-25}$ &$5.34\times10^{-7}$ &159 \\
J0024$-$7204L\tablenotemark{cp} & 50687 -- 54261 & 230.09 & $6.5\times10^{-15}$ &4.9 & $6.55\times10^{-28}$ &$5.82\times10^{-26}$ &$1.27\times10^{-6}$ &89 \\
J0024$-$7204M\tablenotemark{cp} & 48495 -- 54261 & 271.99 & $2.8\times10^{-15}$ &4.9 & $6.55\times10^{-28}$ &$6.14\times10^{-26}$ &$9.61\times10^{-7}$ &94 \\
J0024$-$7204N\tablenotemark{cp} & 48516 -- 54261 & 327.44 & $2.4\times10^{-15}$ &4.9 & $6.55\times10^{-28}$ &$8.35\times10^{-26}$ &$9.02\times10^{-7}$ &128 \\
J0024$-$7204Q\tablenotemark{bcp} & 50690 -- 54261 & 247.94 & $-2.1\times10^{-15}$ &4.9 & $6.55\times10^{-28}$ &$5.74\times10^{-26}$ &$1.08\times10^{-6}$ &88 \\
J0024$-$7204R\tablenotemark{bcp} & 50743 -- 54261 & 287.32 & $-1.2\times10^{-14}$ &4.9 & $6.55\times10^{-28}$ &$5.53\times10^{-26}$ &$7.76\times10^{-7}$ &84 \\
J0024$-$7204S\tablenotemark{bcp} & 50687 -- 54241 & 353.31 & $1.5\times10^{-14}$ &4.9 & $6.55\times10^{-28}$ &$6.82\times10^{-26}$ &$6.33\times10^{-7}$ &104 \\
J0024$-$7204T\tablenotemark{bcp} & 50684 -- 54261 & 131.78 & $-5.1\times10^{-15}$ &4.9 & $6.55\times10^{-28}$ &$3.34\times10^{-26}$ &$2.23\times10^{-6}$ &51 \\
J0024$-$7204U\tablenotemark{bcp} & 48516 -- 54261 & 230.26 & $-5.0\times10^{-15}$ &4.9 & $6.55\times10^{-28}$ &$5.63\times10^{-26}$ &$1.23\times10^{-6}$ &86 \\
J0024$-$7204Y\tablenotemark{bcp} & 51504 -- 54261 & 455.24 & $7.3\times10^{-15}$ &4.9 & $6.55\times10^{-28}$ &$9.42\times10^{-26}$ &$5.26\times10^{-7}$ &144 \\
J0218+4232\tablenotemark{bj} & 49092 -- 54520 & 430.46 & $-1.4\times10^{-14}$\tablenotemark{\dagger} &5.8 & $7.91\times10^{-28}$ &$1.47\times10^{-25}$ &$1.10\times10^{-6}$ &186 \\
J0407+1607\tablenotemark{bj} & 52719 -- 54512 & 38.91 & $-1.2\times10^{-16}$ &4.1 & $3.48\times10^{-28}$ &$6.18\times10^{-26}$ &$3.93\times10^{-5}$ &178 \\
J0437$-$4715\tablenotemark{bp} & 53683 -- 54388 & 173.69 & $-4.7\times10^{-16}$\tablenotemark{\dagger} &0.1 & $8.82\times10^{-27}$ &$5.73\times10^{-25}$ &$6.74\times10^{-7}$ &65 \\
J0613$-$0200\tablenotemark{bjp} & 53406 -- 54520 & 326.60 & $-9.8\times10^{-16}$\tablenotemark{\dagger} &0.5 & $2.91\times10^{-27}$ &$1.11\times10^{-25}$ &$1.18\times10^{-7}$ &38 \\
J0621+1002\tablenotemark{bj} & 52571 -- 54516 & 34.66 & $-5.5\times10^{-17}$\tablenotemark{\dagger} &1.9 & $5.40\times10^{-28}$ &$1.53\times10^{-25}$ &$5.65\times10^{-5}$ &284 \\
J0711$-$6830\tablenotemark{p} & 53687 -- 54388 & 182.12 & $-2.7\times10^{-16}$\tablenotemark{\dagger} &1.0 & $9.52\times10^{-28}$ &$5.00\times10^{-26}$ &$3.71\times10^{-7}$ &53 \\
J0737$-$3039A\tablenotemark{bj} & 53595 -- 54515 & 44.05 & $-3.4\times10^{-15}$\tablenotemark{\dagger} &1.1 & $6.17\times10^{-27}$ &$7.87\times10^{-26}$ &$1.10\times10^{-5}$ &13 \\
J0751+1807\tablenotemark{bj} & 53405 -- 54529 & 287.46 & $-6.3\times10^{-16}$\tablenotemark{\dagger} &0.6 & $1.92\times10^{-27}$ &$1.64\times10^{-25}$ &$2.91\times10^{-7}$ &85 \\
J1012+5307\tablenotemark{bj} & 53403 -- 54523 & 190.27 & $-4.7\times10^{-16}$\tablenotemark{\dagger} &0.5 & $2.43\times10^{-27}$ &$6.94\times10^{-26}$ &$2.36\times10^{-7}$ &29 \\
J1022+1001\tablenotemark{bjp} & 53403 -- 54521 & 60.78 & $-1.6\times10^{-16}$ &0.4 & $3.27\times10^{-27}$ &$4.44\times10^{-26}$ &$1.14\times10^{-6}$ &14 \\
J1024$-$0719\tablenotemark{jp} & 53403 -- 54501 & 193.72 & $-6.9\times10^{-16}$ &0.5 & $2.88\times10^{-27}$ &$5.01\times10^{-26}$ &$1.67\times10^{-7}$ &17 \\
J1045$-$4509\tablenotemark{bp} & 53688 -- 54386 & 133.79 & $-2.0\times10^{-16}$\tablenotemark{\dagger} &3.2 & $3.00\times10^{-28}$ &$4.37\times10^{-26}$ &$1.87\times10^{-6}$ &145 \\
J1455$-$3330\tablenotemark{bj} & 52688 -- 54524 & 125.20 & $-2.5\times10^{-16}$\tablenotemark{\dagger} &0.7 & $1.53\times10^{-27}$ &$5.15\times10^{-26}$ &$5.75\times10^{-7}$ &34 \\
J1600$-$3053\tablenotemark{bp} & 53688 -- 54386 & 277.94 & $-6.5\times10^{-16}$\tablenotemark{\dagger} &2.7 & $4.62\times10^{-28}$ &$5.57\times10^{-26}$ &$4.55\times10^{-7}$ &121 \\
J1603$-$7202\tablenotemark{bp} & 53688 -- 54385 & 67.38 & $-5.9\times10^{-17}$\tablenotemark{\dagger} &1.6 & $4.62\times10^{-28}$ &$2.32\times10^{-26}$ &$1.98\times10^{-6}$ &50 \\
J1623$-$2631\tablenotemark{bcj} & 53403 -- 54517 & 90.29 & $-5.5\times10^{-15}$ &2.2 & $1.46\times10^{-27}$ &$5.81\times10^{-26}$ &$3.71\times10^{-6}$ &40 \\
J1640+2224\tablenotemark{bj} & 53410 -- 54506 & 316.12 & $-1.6\times10^{-16}$\tablenotemark{\dagger} &1.2 & $4.86\times10^{-28}$ &$6.65\times10^{-26}$ &$1.87\times10^{-7}$ &137 \\
J1643$-$1224\tablenotemark{bjp} & 52570 -- 54517 & 216.37 & $-6.8\times10^{-16}$\tablenotemark{\dagger} &4.9 & $2.94\times10^{-28}$ &$4.35\times10^{-26}$ &$1.07\times10^{-6}$ &148 \\
J1701$-$3006A\tablenotemark{bcp} & 53590 -- 54391 & 190.78 & $4.8\times10^{-15}$ &6.9 & $4.65\times10^{-28}$ &$5.82\times10^{-26}$ &$2.61\times10^{-6}$ &125 \\
J1701$-$3006B\tablenotemark{bcp} & 53650 -- 54391 & 278.25 & $2.7\times10^{-14}$ &6.9 & $4.65\times10^{-28}$ &$7.63\times10^{-26}$ &$1.61\times10^{-6}$ &164 \\
J1701$-$3006C\tablenotemark{bcp} & 53590 -- 54396 & 131.36 & $1.1\times10^{-15}$ &6.9 & $4.65\times10^{-28}$ &$3.52\times10^{-26}$ &$3.32\times10^{-6}$ &76 \\
J1713+0747\tablenotemark{bjp} & 53406 -- 54509 & 218.81 & $-3.8\times10^{-16}$\tablenotemark{\dagger} &1.1 & $9.54\times10^{-28}$ &$4.44\times10^{-26}$ &$2.45\times10^{-7}$ &47 \\
J1730$-$2304\tablenotemark{jp} & 52571 -- 54519 & 123.11 & $-3.1\times10^{-16}$ &0.5 & $2.49\times10^{-27}$ &$5.93\times10^{-26}$ &$4.72\times10^{-7}$ &24 \\
J1732$-$5049\tablenotemark{bp} & 53725 -- 54386 & 188.23 & $-4.9\times10^{-16}$ &1.8 & $7.18\times10^{-28}$ &$5.25\times10^{-26}$ &$6.34\times10^{-7}$ &73 \\
J1744$-$1134\tablenotemark{jp} & 52604 -- 54519 & 245.43 & $-4.1\times10^{-16}$\tablenotemark{\dagger} &0.5 & $2.18\times10^{-27}$ &$1.10\times10^{-25}$ &$2.07\times10^{-7}$ &50 \\
J1748$-$2446A\tablenotemark{bcjg} & 52320 -- 54453 & 86.48 & $2.2\times10^{-16}$ &5.5 & $5.83\times10^{-28}$ &$3.89\times10^{-26}$ &$6.77\times10^{-6}$ &67 \\
J1748$-$2446C\tablenotemark{cjg} & 53403 -- 54516 & 118.54 & $8.5\times10^{-15}$ &5.5 & $5.83\times10^{-28}$ &$5.00\times10^{-26}$ &$4.63\times10^{-6}$ &86 \\
J1748$-$2446D\tablenotemark{cg} & 50851 -- 53820 & 212.13 & $-5.7\times10^{-15}$ &5.5 & $5.83\times10^{-28}$ &$6.78\times10^{-26}$ &$1.96\times10^{-6}$ &116 \\
J1748$-$2446E\tablenotemark{bcg} & 53193 -- 53820 & 455.00 & $3.8\times10^{-15}$ &5.5 & $5.83\times10^{-28}$ &$8.95\times10^{-26}$ &$5.62\times10^{-7}$ &153 \\
J1748$-$2446F\tablenotemark{cg} & 53193 -- 53820 & 180.50 & $-1.3\times10^{-16}$ &5.5 & $5.83\times10^{-28}$ &$8.37\times10^{-26}$ &$3.34\times10^{-6}$ &143 \\
J1748$-$2446G\tablenotemark{cg} & 51884 -- 53820 & 46.14 & $-8.4\times10^{-16}$ &5.5 & $5.83\times10^{-28}$ &$5.82\times10^{-26}$ &$3.56\times10^{-5}$ &100 \\
J1748$-$2446H\tablenotemark{cg} & 51884 -- 53820 & 203.01 & $3.4\times10^{-15}$ &5.5 & $5.83\times10^{-28}$ &$7.81\times10^{-26}$ &$2.46\times10^{-6}$ &134 \\
J1748$-$2446I\tablenotemark{bcg} & 50851 -- 54195 & 104.49 & $7.3\times10^{-16}$ &5.5 & $5.83\times10^{-28}$ &$3.54\times10^{-26}$ &$4.21\times10^{-6}$ &61 \\
J1748$-$2446K\tablenotemark{cg} & 51884 -- 53820 & 336.74 & $1.1\times10^{-14}$ &5.5 & $5.83\times10^{-28}$ &$6.67\times10^{-26}$ &$7.65\times10^{-7}$ &114 \\
J1748$-$2446L\tablenotemark{cg} & 51884 -- 53820 & 445.49 & $3.4\times10^{-15}$ &5.5 & $5.83\times10^{-28}$ &$1.39\times10^{-25}$ &$9.09\times10^{-7}$ &238 \\
J1748$-$2446M\tablenotemark{bcg} & 51884 -- 53820 & 280.15 & $-3.9\times10^{-14}$ &5.5 & $5.83\times10^{-28}$ &$1.01\times10^{-25}$ &$1.68\times10^{-6}$ &173 \\
J1748$-$2446N\tablenotemark{bcg} & 53193 -- 54195 & 115.38 & $-7.4\times10^{-15}$ &5.5 & $5.83\times10^{-28}$ &$5.85\times10^{-26}$ &$5.71\times10^{-6}$ &100 \\
J1748$-$2446O\tablenotemark{bcg} & 52500 -- 53957 & 596.43 & $2.5\times10^{-14}$ &5.5 & $5.83\times10^{-28}$ &$2.65\times10^{-25}$ &$9.68\times10^{-7}$ &454 \\
J1748$-$2446P\tablenotemark{bcg} & 53193 -- 54557 & 578.50 & $-8.7\times10^{-14}$ &5.5 & $5.83\times10^{-28}$ &$1.56\times10^{-25}$ &$6.08\times10^{-7}$ &267 \\
J1748$-$2446Q\tablenotemark{bcg} & 53193 -- 54139 & 355.62 & $4.6\times10^{-15}$ &5.5 & $5.83\times10^{-28}$ &$8.80\times10^{-26}$ &$9.05\times10^{-7}$ &151 \\
J1748$-$2446R\tablenotemark{cg} & 52500 -- 53820 & 198.86 & $-1.9\times10^{-14}$ &5.5 & $5.83\times10^{-28}$ &$8.23\times10^{-26}$ &$2.71\times10^{-6}$ &141 \\
J1748$-$2446S\tablenotemark{cg} & 53193 -- 53820 & 163.49 & $-1.7\times10^{-15}$ &5.5 & $5.83\times10^{-28}$ &$4.46\times10^{-26}$ &$2.17\times10^{-6}$ &76 \\
J1748$-$2446T\tablenotemark{cg} & 51884 -- 53819 & 141.15 & $-6.1\times10^{-15}$ &5.5 & $5.83\times10^{-28}$ &$5.12\times10^{-26}$ &$3.34\times10^{-6}$ &88 \\
J1748$-$2446V\tablenotemark{bcg} & 53193 -- 53820 & 482.51 & $2.2\times10^{-14}$ &5.5 & $5.83\times10^{-28}$ &$1.26\times10^{-25}$ &$7.04\times10^{-7}$ &216 \\
J1748$-$2446W\tablenotemark{bcg} & 52500 -- 53820 & 237.80 & $-7.1\times10^{-15}$ &5.5 & $5.83\times10^{-28}$ &$9.57\times10^{-26}$ &$2.20\times10^{-6}$ &164 \\
J1748$-$2446X\tablenotemark{bcg} & 51884 -- 54139 & 333.44 & $-6.5\times10^{-15}$ &5.5 & $5.83\times10^{-28}$ &$8.18\times10^{-26}$ &$9.57\times10^{-7}$ &140 \\
J1748$-$2446Y\tablenotemark{bcg} & 53193 -- 53820 & 488.24 & $-4.0\times10^{-14}$ &5.5 & $5.83\times10^{-28}$ &$2.10\times10^{-25}$ &$1.15\times10^{-6}$ &360 \\
J1748$-$2446Z\tablenotemark{bcg} & 53193 -- 54139 & 406.08 & $1.4\times10^{-14}$ &5.5 & $5.83\times10^{-28}$ &$8.43\times10^{-26}$ &$6.65\times10^{-7}$ &145 \\
J1748$-$2446aa\tablenotemark{cg} & 51884 -- 53819 & 172.77 & $1.3\times10^{-14}$ &5.5 & $5.83\times10^{-28}$ &$2.28\times10^{-25}$ &$9.92\times10^{-6}$ &391 \\
J1748$-$2446ab\tablenotemark{cg} & 51884 -- 53819 & 195.32 & $-1.6\times10^{-14}$ &5.5 & $5.83\times10^{-28}$ &$4.67\times10^{-26}$ &$1.59\times10^{-6}$ &80 \\
J1748$-$2446ac\tablenotemark{cg} & 52500 -- 53819 & 196.58 & $-8.8\times10^{-15}$ &5.5 & $5.83\times10^{-28}$ &$7.19\times10^{-26}$ &$2.42\times10^{-6}$ &123 \\
J1748$-$2446ad\tablenotemark{bcg} & 53204 -- 54557 & 716.36 & $1.7\times10^{-14}$ &5.5 & $5.83\times10^{-28}$ &$1.77\times10^{-25}$ &$4.48\times10^{-7}$ &303 \\
J1748$-$2446ae\tablenotemark{bcg} & 53193 -- 53820 & 273.33 & $4.3\times10^{-14}$ &5.5 & $5.83\times10^{-28}$ &$6.57\times10^{-26}$ &$1.14\times10^{-6}$ &113 \\
J1748$-$2446af\tablenotemark{cg} & 53193 -- 53820 & 302.63 & $2.1\times10^{-14}$ &5.5 & $5.83\times10^{-28}$ &$1.07\times10^{-25}$ &$1.52\times10^{-6}$ &183 \\
J1748$-$2446ag\tablenotemark{cg} & 53193 -- 53819 & 224.82 & $-6.3\times10^{-16}$ &5.5 & $5.83\times10^{-28}$ &$9.49\times10^{-26}$ &$2.44\times10^{-6}$ &163 \\
J1748$-$2446ah\tablenotemark{cg} & 53193 -- 53819 & 201.40 & $-2.3\times10^{-14}$ &5.5 & $5.83\times10^{-28}$ &$5.49\times10^{-26}$ &$1.76\times10^{-6}$ &94 \\
J1756$-$2251\tablenotemark{bj} & 53403 -- 54530 & 35.14 & $-1.3\times10^{-15}$ &2.9 & $1.65\times10^{-27}$ &$9.70\times10^{-26}$ &$5.42\times10^{-5}$ &59 \\
J1801$-$1417\tablenotemark{j} & 53405 -- 54505 & 275.85 & $-4.0\times10^{-16}$ &1.8 & $5.42\times10^{-28}$ &$6.15\times10^{-26}$ &$3.44\times10^{-7}$ &113 \\
J1803$-$30\tablenotemark{cp} & 53654 -- 54379 & 140.83 & $-1.0\times10^{-15}$ &7.8 & $4.11\times10^{-28}$ &$5.51\times10^{-26}$ &$5.12\times10^{-6}$ &134 \\
J1804$-$0735\tablenotemark{bcj} & 52573 -- 54518 & 43.29 & $-8.8\times10^{-16}$ &8.4 & $3.82\times10^{-28}$ &$8.44\times10^{-26}$ &$8.95\times10^{-5}$ &221 \\
J1804$-$2717\tablenotemark{bj} & 52574 -- 54453 & 107.03 & $-4.7\times10^{-16}$ &1.2 & $1.44\times10^{-27}$ &$2.40\times10^{-26}$ &$5.79\times10^{-7}$ &17 \\
J1807$-$2459A\tablenotemark{bcp} & 53621 -- 54462 & 326.86 & $4.8\times10^{-16}$ &2.7 & $1.19\times10^{-27}$ &$1.53\times10^{-25}$ &$9.13\times10^{-7}$ &129 \\
J1810$-$2005\tablenotemark{bj} & 53406 -- 54508 & 30.47 & $-1.4\times10^{-16}$ &4.0 & $4.28\times10^{-28}$ &$2.22\times10^{-25}$ &$2.28\times10^{-4}$ &519 \\
J1823$-$3021A\tablenotemark{cj} & 53403 -- 54530 & 183.82 & $-1.1\times10^{-13}$ &7.9 & $4.06\times10^{-28}$ &$3.93\times10^{-26}$ &$2.17\times10^{-6}$ &97 \\
J1824$-$2452A\tablenotemark{cjpg} & 53403 -- 54509 & 327.41 & $-1.7\times10^{-13}$ &4.9 & $3.79\times10^{-27}$ &$7.80\times10^{-26}$ &$8.43\times10^{-7}$ &21 \\
J1824$-$2452B\tablenotemark{cg} & 53629 -- 54201 & 152.75 & $5.6\times10^{-15}$ &4.9 & $6.55\times10^{-28}$ &$4.26\times10^{-26}$ &$2.11\times10^{-6}$ &65 \\
J1824$-$2452C\tablenotemark{bcg} & 52335 -- 54202 & 240.48 & $-9.8\times10^{-15}$ &4.9 & $6.55\times10^{-28}$ &$6.48\times10^{-26}$ &$1.30\times10^{-6}$ &99 \\
J1824$-$2452E\tablenotemark{cg} & 53629 -- 54201 & 184.53 & $3.7\times10^{-15}$ &4.9 & $6.55\times10^{-28}$ &$7.51\times10^{-26}$ &$2.55\times10^{-6}$ &115 \\
J1824$-$2452F\tablenotemark{cg} & 52497 -- 54114 & 407.97 & $-1.6\times10^{-15}$ &4.9 & $6.55\times10^{-28}$ &$9.74\times10^{-26}$ &$6.78\times10^{-7}$ &149 \\
J1824$-$2452G\tablenotemark{bcg} & 53629 -- 54202 & 169.23 & $-5.2\times10^{-15}$ &4.9 & $6.55\times10^{-28}$ &$7.23\times10^{-26}$ &$2.93\times10^{-6}$ &110 \\
J1824$-$2452H\tablenotemark{bcg} & 53629 -- 54202 & 216.01 & $-3.6\times10^{-15}$ &4.9 & $6.55\times10^{-28}$ &$8.27\times10^{-26}$ &$2.05\times10^{-6}$ &126 \\
J1824$-$2452J\tablenotemark{bcg} & 53629 -- 54201 & 247.54 & $4.7\times10^{-15}$ &4.9 & $6.55\times10^{-28}$ &$1.07\times10^{-25}$ &$2.03\times10^{-6}$ &163 \\
J1841+0130\tablenotemark{bj} & 53405 -- 54513 & 33.59 & $-9.2\times10^{-15}$ &3.2 & $4.19\times10^{-27}$ &$1.65\times10^{-25}$ &$1.10\times10^{-4}$ &39 \\
J1843$-$1113\tablenotemark{j} & 53353 -- 54508 & 541.81 & $-2.8\times10^{-15}$ &2.0 & $9.33\times10^{-28}$ &$1.64\times10^{-25}$ &$2.61\times10^{-7}$ &176 \\
J1857+0943\tablenotemark{bjp} & 53409 -- 54517 & 186.49 & $-6.0\times10^{-16}$\tablenotemark{\dagger} &0.9 & $1.59\times10^{-27}$ &$7.27\times10^{-26}$ &$4.50\times10^{-7}$ &46 \\
J1905+0400\tablenotemark{j} & 53407 -- 54512 & 264.24 & $-3.4\times10^{-16}$ &1.3 & $6.81\times10^{-28}$ &$7.40\times10^{-26}$ &$3.36\times10^{-7}$ &109 \\
J1909$-$3744\tablenotemark{bp} & 53687 -- 54388 & 339.32 & $-3.1\times10^{-16}$\tablenotemark{\dagger} &1.1 & $6.78\times10^{-28}$ &$8.09\times10^{-26}$ &$1.89\times10^{-7}$ &119 \\
J1910$-$5959A\tablenotemark{bcp} & 53666 -- 54380 & 306.17 & $-2.8\times10^{-16}$ &4.5 & $7.13\times10^{-28}$ &$7.71\times10^{-26}$ &$8.75\times10^{-7}$ &108 \\
J1910$-$5959B\tablenotemark{cp} & 53609 -- 54473 & 119.65 & $1.1\times10^{-14}$ &4.5 & $7.13\times10^{-28}$ &$3.81\times10^{-26}$ &$2.83\times10^{-6}$ &53 \\
J1910$-$5959C\tablenotemark{cp} & 53666 -- 54390 & 189.49 & $-7.8\times10^{-17}$ &4.5 & $7.13\times10^{-28}$ &$4.34\times10^{-26}$ &$1.29\times10^{-6}$ &61 \\
J1910$-$5959D\tablenotemark{cp} & 53621 -- 54460 & 110.68 & $-1.2\times10^{-14}$ &4.5 & $7.13\times10^{-28}$ &$3.03\times10^{-26}$ &$2.63\times10^{-6}$ &42 \\
J1910$-$5959E\tablenotemark{cp} & 53610 -- 54441 & 218.73 & $2.1\times10^{-14}$ &4.5 & $7.13\times10^{-28}$ &$4.77\times10^{-26}$ &$1.06\times10^{-6}$ &67 \\
J1911+1347\tablenotemark{j} & 53403 -- 54530 & 216.17 & $-8.0\times10^{-16}$ &1.6 & $9.63\times10^{-28}$ &$7.00\times10^{-26}$ &$5.70\times10^{-7}$ &73 \\
J1911$-$1114\tablenotemark{bj} & 53407 -- 54512 & 275.81 & $-4.8\times10^{-16}$\tablenotemark{\dagger} &1.6 & $6.66\times10^{-28}$ &$5.62\times10^{-26}$ &$2.78\times10^{-7}$ &84 \\
J1913+1011\tablenotemark{j} & 53745 -- 54911 & 27.85 & $-2.6\times10^{-12}$ &4.5 & $5.51\times10^{-26}$ &$2.14\times10^{-25}$ &$2.93\times10^{-4}$ &3.9 \\
J1939+2134\tablenotemark{jp} & 53407 -- 54519 & 641.93 & $-4.3\times10^{-14}$\tablenotemark{\dagger} &3.5 & $1.86\times10^{-27}$ &$1.79\times10^{-25}$ &$3.65\times10^{-7}$ &96 \\
J1955+2908\tablenotemark{bj} & 53403 -- 54524 & 163.05 & $-7.6\times10^{-16}$\tablenotemark{\dagger} &5.4 & $3.23\times10^{-28}$ &$7.07\times10^{-26}$ &$3.39\times10^{-6}$ &219 \\
J2019+2425\tablenotemark{bj} & 53599 -- 54505 & 254.16 & $-1.7\times10^{-16}$\tablenotemark{\dagger} &0.9 & $7.14\times10^{-28}$ &$9.23\times10^{-26}$ &$3.07\times10^{-7}$ &129 \\
J2033+17\tablenotemark{bj} & 53702 -- 54522 & 168.10 & $-3.1\times10^{-16}$ &1.4 & $7.93\times10^{-28}$ &$7.49\times10^{-26}$ &$8.65\times10^{-7}$ &94 \\
J2051$-$0827\tablenotemark{bj} & 53410 -- 54520 & 221.80 & $-6.1\times10^{-16}$\tablenotemark{\dagger} &1.3 & $1.04\times10^{-27}$ &$7.57\times10^{-26}$ &$4.65\times10^{-7}$ &73 \\
J2124$-$3358\tablenotemark{jp} & 53410 -- 54510 & 202.79 & $-5.1\times10^{-16}$\tablenotemark{\dagger} &0.2 & $5.13\times10^{-27}$ &$4.85\times10^{-26}$ &$6.96\times10^{-8}$ &9.4 \\
J2129$-$5721\tablenotemark{bp} & 53687 -- 54388 & 268.36 & $-2.0\times10^{-15}$\tablenotemark{\dagger} &2.5 & $8.71\times10^{-28}$ &$6.12\times10^{-26}$ &$5.13\times10^{-7}$ &70 \\
J2145$-$0750\tablenotemark{bjp} & 53409 -- 54510 & 62.30 & $-1.0\times10^{-16}$\tablenotemark{\dagger} &0.5 & $2.05\times10^{-27}$ &$3.83\times10^{-26}$ &$1.17\times10^{-6}$ &19 \\
J2229+2643\tablenotemark{bj} & 53403 -- 54524 & 335.82 & $-1.6\times10^{-16}$ &1.4 & $3.95\times10^{-28}$ &$9.89\times10^{-26}$ &$2.96\times10^{-7}$ &250 \\
J2317+1439\tablenotemark{bj} & 53406 -- 54520 & 290.25 & $-1.3\times10^{-16}$\tablenotemark{\dagger} &1.9 & $2.82\times10^{-28}$ &$8.83\times10^{-26}$ &$4.68\times10^{-7}$ &313 \\
J2322+2057\tablenotemark{j} & 53404 -- 54519 & 207.97 & $-1.8\times10^{-16}$\tablenotemark{\dagger} &0.8 & $9.55\times10^{-28}$ &$1.12\times10^{-25}$ &$4.78\times10^{-7}$ &117 \\